\documentclass[fleqn,usenatbib]{mnras}

\usepackage{graphicx}
\usepackage{subfigure}
\usepackage{verbatim}
\usepackage{color}
\usepackage[figuresright]{rotating}
\usepackage{longtable}
\usepackage{hyperref}

\usepackage[T1]{fontenc}

\DeclareRobustCommand{\VAN}[3]{#2}
\let\VANthebibliography\thebibliography
\def\thebibliography{\DeclareRobustCommand{\VAN}[3]{##3}\VANthebibliography}

\usepackage{graphicx}	
\usepackage{amsmath}	
\usepackage{amssymb}	

\title[A Spectral data release]{A spectral data release for 104 Type II Supernovae from the Tsinghua Supernova Group}

\author[Lin et al.]{
Han Lin,$^{1,2,3,5}$ 
Xiaofeng Wang,$^{1}$\thanks{E-mail:wang\_xf@mail.tsinghua.edu.cn}
Jujia Zhang,$^{3,4,5}$\thanks{E-mail:jujia@ynao.ac.cn}Shengyu Yan,$^1$
Danfeng Xiang,$^1$
Tianmeng Zhang,$^{6,7,8}$
\newauthor
Xulin Zhao,$^{9}$
Xinghan Zhang,$^{10}$
Hanna Sai,$^1$
Liming Rui,$^1$
Jun Mo,$^1$ 
Gaobo Xi,$^1$
Fang Huang,$^{11}$
Xue Li,$^1$
\newauthor
Yongzhi Cai,$^{1,3,4,5}$
Weili Lin,$^1$
Jie Lin,$^{1,12}$
Chengyuan Wu,$^{3,1}$
Jicheng Zhang,$^{13,1}$
Zhihao Chen,$^1$
Zhitong Li,$^1$
\newauthor
Wenxiong Li,$^{14}$
Linyi Li,$^1$
Kaicheng Zhang,$^1$
Cheng Miao,$^1$
Juncheng Chen,$^{15}$
Zhou Fan ,$^{14}$
Jianning Fu,$^{13}$
\newauthor
Shengbang Qian,$^{16}$
Hong Wu,$^{17}$
Xue-Bing Wu,$^{18}$
Jingzhi Yan,$^{19}$
Huawei Zhang,$^{18}$
Junbo Zhang,$^{17}$
\newauthor
Liyun Zhang,$^{20}$
Jie Zheng,$^{14}$
Qian Zhai$^{3}$
\\
$^{1}$Physics Department, Tsinghua University, Beijing, 100084, China\\
$^{2}$Key Laboratory of Radio Astronomy and Technology, Chinese Academy of Sciences, A20 Datun Road, Chaoyang District, Beijing, 100101, P. R. China\\
$^{3}$Yunnan Observatories, Chinese Academy of Sciences, Kunming 650011, China\\
$^{4}$Key Laboratory for the Structure and Evolution of Celestial Objects, Chinese Academy of Sciences, Kunming 650011, China\\
$^{5}$International Centre of Supernovae, Yunnan Key Laboratory, Kunming 650216, P. R. China\\
$^{6}$Institute for Frontiers in Astronomy and Astrophysics, Beijing Normal University, Beijing, 102206, China\\
$^{7}$Key Laboratory of Space Astronomy and Technology, National Astronomical Observatories, Chinese Academy of Sciences, 20A Datun Road, Beijing 100101, China\\
$^{8}$School of Astronomy and Space Science, University of Chinese Academy of Sciences, Beijing 101408, China\\
$^{9}$School of Science, Tianjin University of Technology, Tianjin 300384, China\\
$^{10}$ School of Physics and Information Engineering, Jiangsu Second Normal University, Nanjing, Jiangsu 211200, PR China\\
$^{11}$Department of Astronomy, Shanghai Jiao Tong University, Shanghai 200240, China\\
$^{12}$Department of Astronomy, University of Science and Technology, Hefei, China \\
$^{13}$Department of Astronomy, Beijing Normal University, Beijing, 1000875, China\\
$^{14}$Key Laboratory of Optical Astronomy, National Astronomical Observatories, Chinese Academy of Sciences, Beijing 100101, China\\
$^{15}$Wuzhou University \\  
$^{16}$Department of Astronomy, School of Physics and Astronomy, Yunnan University, Kunming 650091, P. R. China\\
$^{17}$National Astronomical Observatories, Chinese Academy of Sciences, Beijing 100101, China\\
$^{18}$Department of Astronomy, School of Physics, Peking University, Beijing 100871, China\\
$^{19}$Purple Mountain Observatory, Chinese Academy of Sciences, Nanjing, China\\
$^{20}$College of Physics, Guizhou University, Guiyang, China\\
}

\date{Accepted XXX. Received YYY; in original form ZZZ}

\pubyear{2015}

\begin{document}
\label{firstpage}
\pagerange{\pageref{firstpage}--\pageref{lastpage}}
\maketitle

\begin{abstract}
We present 206 unpublished optical spectra of 104 type II supernovae obtained by the Xinglong 2.16m telescope and Lijiang 2.4m telescope during the period from 2011 to 2018, spanning the phases from about 1 to 200 days after the SN explosion. 
The spectral line identifications, evolution of line velocities and pseudo equivalent widths, as well as correlations between some important spectral parameters are presented. 
Our sample displays a large range in expansion velocities. For instance, the Fe~{\sc ii} $5169$ velocities measured from spectra at $t\sim 50$ days after the explosion vary from ${\rm 2000\ km\ s^{-1}}$ to ${\rm 5500\ km\ s^{-1}}$, with an average value of ${\rm 3872 \pm 949\ km\ s^{-1}}$. 
Power-law functions can be used to fit the velocity evolution, with the power-law exponent quantifying the velocity decline rate.
We found an anticorrelation existing between H$\beta$ velocity at mid-plateau phase and its velocity decay exponent, SNe II with higher velocities tending to have smaller velocity decay rate. 
Moreover, we noticed that the velocity decay rate inferred from the Balmer lines (i.e., H$\alpha$ and H$\beta$) have moderate correlations with the ratio of absorption to emission for H$\alpha$ (a/e).
In our sample, two objects show possibly flash-ionized features at early phases.
Besides, we noticed that multiple high-velocity components may exist on the blue side of hydrogen lines of SN 2013ab, possibly suggesting that these features arise from complex line forming region.
All our spectra can be found in \href{https://www.wiserep.org}{WISeREP} and \href{https://doi.org/10.5281/zenodo.10466160}{Zenodo}.

\end{abstract}

\begin{keywords}
Techniques: spectroscopic -- surveys -- supernovae:general
\end{keywords}



\section{Introduction}

Supernovae (SNe) with prominent hydrogen lines in their optical spectra are classified as type II supernovae (SNe II) \citep{1941PASP...53..224M,1997ARA&A..35..309F}. 
It is generally accepted that SNe II are produced by explosion of massive stars with initial masses $ \geq 8 {\rm M}\odot$ \citep{2003PASP..115.1289V,2009MNRAS.395.1409S,2009ARA&A..47...63S}.
The characteristic of SNe II display a wide variety in photometric and spectroscopic features. Based on the spectral and photometric features, SNe II can be divided into subclasses of SNe IIP, IIL, IIb, and IIn etc. 
The subclasses of SNe IIP and IIL are primarily based on the shape of light curves. 
Those with almost constant luminosity (plateau) in light curves are called type IIP while those with a linear decline after maximun light are called type IIL \citep{1979A&A....72..287B}.  
Direct identifications of the progenitors of nearby SNe IIP, e.g. 
SN 2005cs \citep{2005MNRAS.364L..33M,2006ApJ...641.1060L}, SN 2012aw \citep{2012ApJ...756..131V,2012ApJ...759L..13F}, SN 2017eaw \citep{2018MNRAS.481.2536K,2019ApJ...875..136V,2020MNRAS.494.5882R,2019MNRAS.485.1990R}, et al. suggested that SNe IIP arise from red supergiant stars (RSG) with initial masses of 8-17 ${\rm M}\odot$.
Among SNe IIP, some have low expansion velocities and less amount of nickel synthesized in the explosion, which are thought to represent the low luminosity tail of a continuous distribution in the parameter space of SNe IIP and probably originate from intermediate-mass ($10-15{\rm M\odot}$) stars \citep{2014MNRAS.439.2873S,2014ApJ...797....5Z}.
Although SNe IIP and SNe IIL have different lightcurve shapes, there is increasing studies showing that no significant distinction exists between the observational properties of these two subclasses  \citep{2014ApJ...786...67A,2015ApJ...799..208S,2016MNRAS.459.3939V,2016AJ....151...33G,2019MNRAS.490.2799D}.  \citet{2014ApJ...786...67A} found that 
more luminous SNe II tend to have faster post-peak declines and this trend could be used to estimate distances to SNe II (i.e., the photometric color method, \citealt{2015ApJ...815..121D}).
\citet{2014ApJ...786L..15G} also suggested that there is no definitive spectral distinction between SNe IIP and SNe IIL. More luminous SNe II tend to have larger H$\alpha$ velocities and smaller ratios of absorption
to emission (a/e) of H$\alpha$. 
Theoritical studies show that the plateaus in SNe IIP result from the combination of hydrogen envelope \citep{1983Ap&SS..89...89L,1992SvAL...18...43B,1993ApJ...414..712P,2015ApJ...814...63M,2016MNRAS.455..423M}, and the formation of SNe IIL can be due to that the exploding stars have less amount of hydrogen envelope \citep{2014ApJ...786...67A,2017ApJ...850...90G}.

Besides SNe IIP and SNe IIL, supernovae with long lasting narrow or/and intermediate-width emission lines of hydrogen in their optical spectra are classified as SNe IIn \citep{1990MNRAS.244..269S}. Different from `normal' type IIP and type IIL SNe, circumstellar interaction (CSI) plays important roles in the observed properties of SNe IIn \citep{2017hsn..book..403S}.
The spectra of some SNe II will evolve like SNe Ib a few weeks after the maximum light, which are called SNe IIb and 
are thought to be the transitional objects linking between SNe II with stripped-envelope SNe Ib \citep{1993ApJ...415L.103F}. 
An even more rare subclass  of SNe II is 1987A-like events, which shows unusually long rise to the peak \citep{1988AJ.....95...63H,2023MNRAS.520.2965X}. We do not include the subclasses of SNe IIn, SNe IIb and 1987A-like events in this work.

With the wide-field, high-cadence transient surveys, the observed properties of both interesting individual objects and large samples of SNe II have been studied. 
Spectroscopic observations provide important information on the physical origins and even stellar winds blown from their progenitors when the spectra can be obtained in very early phases. 
For example, the flash-ionized signatures allow constraints on the density and velocity of circumstellar material (CSM), and therefore the mass loss history shortly before the SN explosion \citep{2014Natur.509..471G,2015ApJ...806..213S,2016ApJ...818....3K,2017NatPh..13..510Y,2021MNRAS.505.4890L,2023SciBu..68.2548Z}. On the other hand, the spectra in the nebular phase can be used to estimate the progenitor mass \citep{2012MNRAS.420.3451M,2012A&A...546A..28J,2017MNRAS.467..369S}. 
In this paper, we present data and analysis of 206 spectra for 104 SNe II obtained by Tsinghua Supernovae Group during the period from 2011 to 2018.
We describe our sample in Section 2. 
The spectral evolution and line identifications are presented 
in Section 3. 
The evolution of expansion velocities and the pseudo equivalent width (pEW) of some important spectral lines are presented in Sections 4 and 5, respectively. 
The correlations between these spectrocopic parameters are given 
in Section 6, and we conclude in Section 7.

\section{Data Sample}

Our sample consists of 206 unpublished spectra for 104 SNe IIP/L observed between 2011 and 2018. The spectra were obtained using the Xinglong 2.16-m telescope (+BFOSC/OMR; \citealt{2016PASP..128k5005F}) of NAOC and the Lijiang 2.4-m telescope (+YFOSC; \citealt{2015RAA....15..918F}) of Yunnan Observatories . All the spectra were reduced using standard IRAF pipelines \citep{1986SPIE..627..733T,1993ASPC...52..173T},  including bias and flat-field correction, cosmic-ray removal, wavelength and flux calibration. All spectra were corrected for atmospheric extinction using the extinction curves of the local observatories. Telluric absorptions were also removed for the spectra whenever possible.

\begin{figure}
	\includegraphics[width=1\columnwidth]{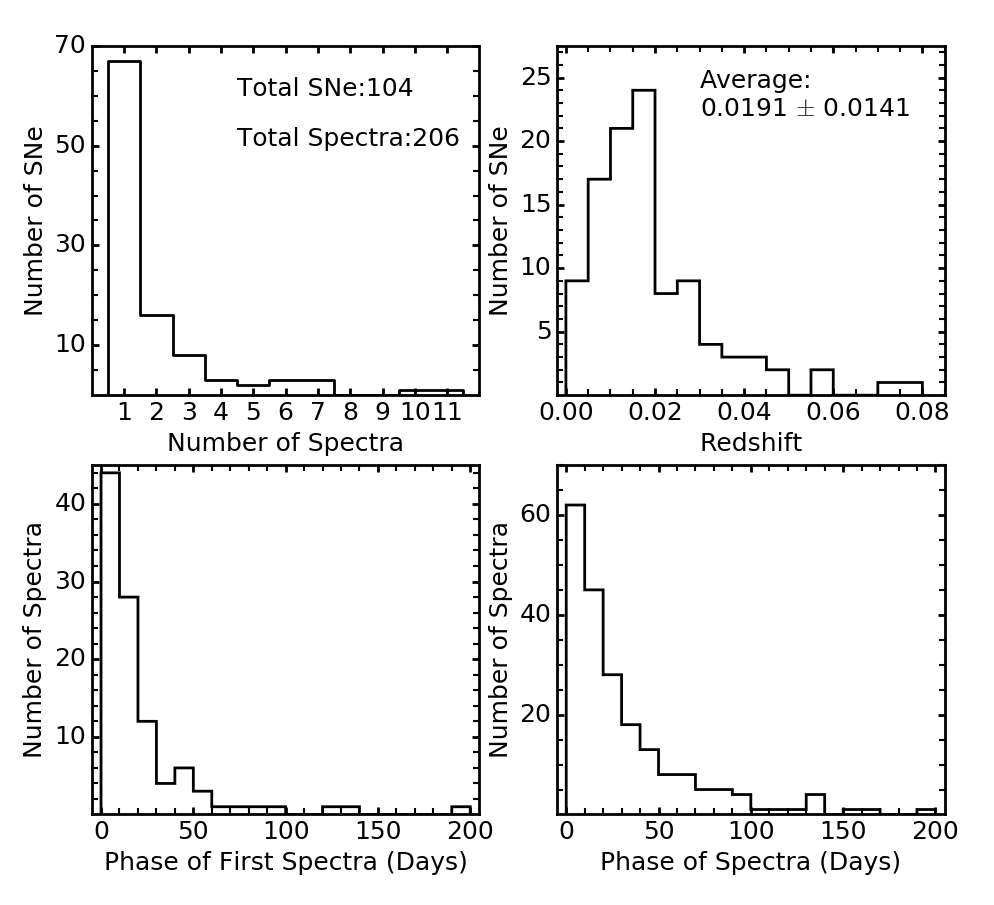}
    \caption{Distribution of SN-level and spectrum-level parameters. The upper-left panel shows the number of spectra per SN. The upper-right panel shows the redshift distribution of our sample. The lower-left and lower-right panels show the distribution of the phase of the first spectrum and all spectra, respectively.}
    \label{fig:histogram}
\end{figure}

For each SN, we used the same method following \citet{2017ApJ...850...89G} to estimate their explosion dates. 
If high-cadence observations are available for the sample near the discovery, we adopt the midpoints between the last non-detection and the first discovery as the estimated explosion date. The lower and upper limits of the estimated explosion date will cover the non-detection and first-discovery epochs. For the object with sparse or without pre-discovery images, the explosion dates were estimated by matching spectral templates via the Supernova Identification (SNID) code \citep{2007ApJ...666.1024B}. SNID automatically gives matches that satisfy the default rlap cut (rlap$_{min} \geq 5$). In general, the best-fit templates are the first several templates that have the largest rlap values. The goodness of matching was also checked by eyes and we chose the best two from those rlap-ordered templates in most cases. The details of spectral matching and the plots with the best matches for each SN in our sample are shown in the supplementary materials. The explosion date is then taken as the average epoch of the two best-fit templates and the uncertainty is taken as the standard deviation of the average value. 
For those that the explosion dates can not be estimated from either the midpoint or the SNID fit, we have to treat the time of first discovery as the explosion dates\footnote{Actually, for these objects, the time of first discovery is the upper limit of explosion date.}. 
Among our 104 objects, 25 (24\%) explosion epochs were obtained using midpoint, 35 (34\%) were obtained using SNID template matching,  while 25 (24\%) have to use the time of first discovery as the estimated explosion time.
It should be noted that some of our SN II samples have already been studied, and 
the explosion dates of this portion (about 18\% of our sample) were taken from the literatures.
The host galaxy redshifts of all our sample were taken from the NASA/IPAC extra galactic Database (NED) and the Galactic extinctions were from \citet{2011ApJ...737..103S}.
The information of each SN II of our sample is listed in Table \ref{table:snlog}.

Figure \ref{fig:histogram} shows the statistical distribution of our sample, including the number of spectra, host-galaxy redshift, the phases of first spectrum and all spectra, respectively. A total of 206 spectra were collected for 104 objects. These spectra cover the phases from 1 days to 200 days after the explosion. Of the 104 SNe IIP/L, 37 objects have at least two spectra, while PS 15cwo has 11 spectra. A total of 20 objects (19\%) were observed earlier than 5 days after the explosion and 46 object (44\%) were observed earlier than 10 days after the explosion. 
The redshifts of these SNe II range from 0.00199 (SN 2012A) to 0.08 (SN 2017hxz), with a median value of 0.016, among which 78 objects (75\%) are at hubble-flow distances (z $>$ 0.01).

\section{Spectral Evolution and Line Identification}

\subsection{Overall evolution of the spectra}

\begin{figure*}
	\includegraphics[width=1.8\columnwidth]{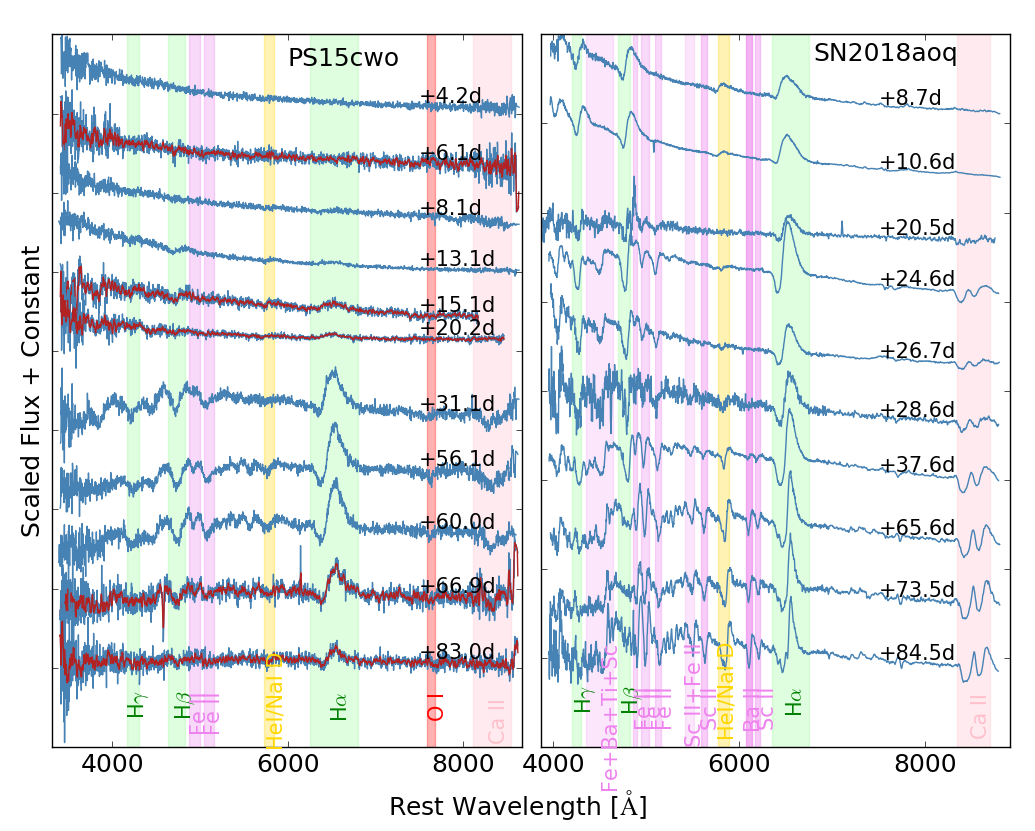}
	\caption{Spectral evolution and line identifications of PS 15cwo and SN 2018aoq. Shaded regions of different colors represent different spectral lines, with green representing hydrogen, yellow representing He I or Na I, purple  representing iron, barium and other metal elements, respectively}
	\label{fig:d2_specline}
\end{figure*}

The spectral evolution of our sample is shown in Figure \ref{fig:sn_spec}.
The evolution of two well-observed SNe IIL (PS15cwo and SN 2018aoq), and the spectral line identifications are shown in Figure \ref{fig:d2_specline}. One can see that the spectra at early phases are characterized by P-Cgyni profile of Balmer series and He~{\sc i} lines. 
The He~{\sc i} 5876 gradually evolves into Na I D absorption as the temperature decreases. 
For example, in the spectra of SN 2018aoq, the P-Cgyni profile near 5850 \AA    $\ $disappeared at $t\sim 20.5$ days after the explosion and 
it reappeared in the $t\sim 24.6$ day spectrum, with the new emergent feature being due to the Na~{\sc i} D absorption.
As the SN ejecta expands and gets cooler, features from the inner materials start to appear in the spectra. 
For instance, at $t\sim 1$ month, the absorption feature of Fe~{\sc ii} lines near 4800 \AA \   to 4900 \AA , including Fe~{\sc ii} $5169, 5018, 4929$, start to appear and get stronger with time. 
Later on, absorption features of other elements such as Sc~{\sc ii}  6247, 5663, Ba~{\sc ii} 6142, Ca~{\sc ii} near-infrared (NIR) triplet and O~{\sc i} $7774$ become visible in the spectra, and absorption features of Ba, Ti and Sc are blended with Fe~{\sc ii} lines. 

\begin{figure}
	\includegraphics[width=1\columnwidth]{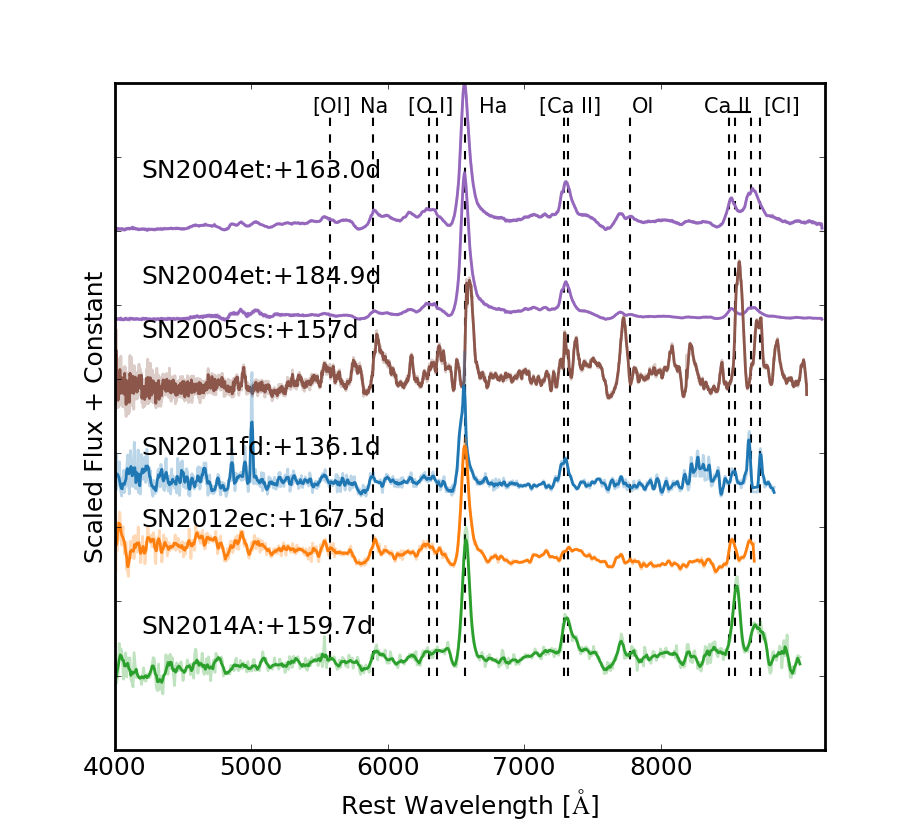}
    \caption{Nebular-phase spectra and line identifications for several SNe II objects in our sample.}
    \label{fig:nebula_plot}
\end{figure}

In the nebular phase, when the SN ejecta became transparent, the supernova is powered by the radioactive decay of ${\rm ^{56}Co}$. At this late phase, the nebular spectra are dominated by emission lines due to recombination, collisional excitation and fluorescence \citep{2001astro.ph.11573B}. 
The prominent features include emission lines of H$\alpha$, [O~{\sc i}] 6300,6343, [Ca~{\sc ii}]  7291,7323 and Ca~{\sc ii} NIR triplet (see Figure \ref{fig:nebula_plot}).

\subsection{High velocity features of hydrogen}
A few of SNe II are found to exhibit an extra absorption component on the blue side of Ha lines. The notch could be the high velocity (HV) feature of H$\alpha$ or absorption of other species. \citet{2017ApJ...850...89G} suggested that this absorption feature observed before t$\sim$35 days after explosion can be associated with Si~{\sc ii} $6355$ line, while those observed after the middle of the plateau are HV features of hydrogen.
There are two explanations proposed for the origin of the HV features. \citet{2000ApJ...545..444B} suggested that the blue side absorption can be due to the second P-Cgyni profile of H$\alpha$ caused by non-LTE effects. 
They reproduced the second absorption of ${\rm H\beta}$ line in their synthetic spectra with the full non-LTE atmosphere code PHOENIX. 
On the other hand, the HV features were also regarded as signs of interaction between SN ejecta and the CSM \citep{2007ApJ...662.1136C}, which can be used to estimate the mass loss rate of the progenitor star. In theory, the X-rays from the reverse shock can ionize and excite the outer layers of the ejecta, producing a shallow depression in the blue wings of absorption part of P-Cgyni profile\citep{2007ApJ...662.1136C}. Moreover, the HV features can also result from cold dense shell, and such an absorption is usually deeper \citep{2007ApJ...662.1136C}.
Owing to the observed diversity of absorption shape, \citet{2017ApJ...850...89G} suggested that these HV features are most likely produced by interaction.

\begin{figure}
	\includegraphics[width=1\columnwidth]{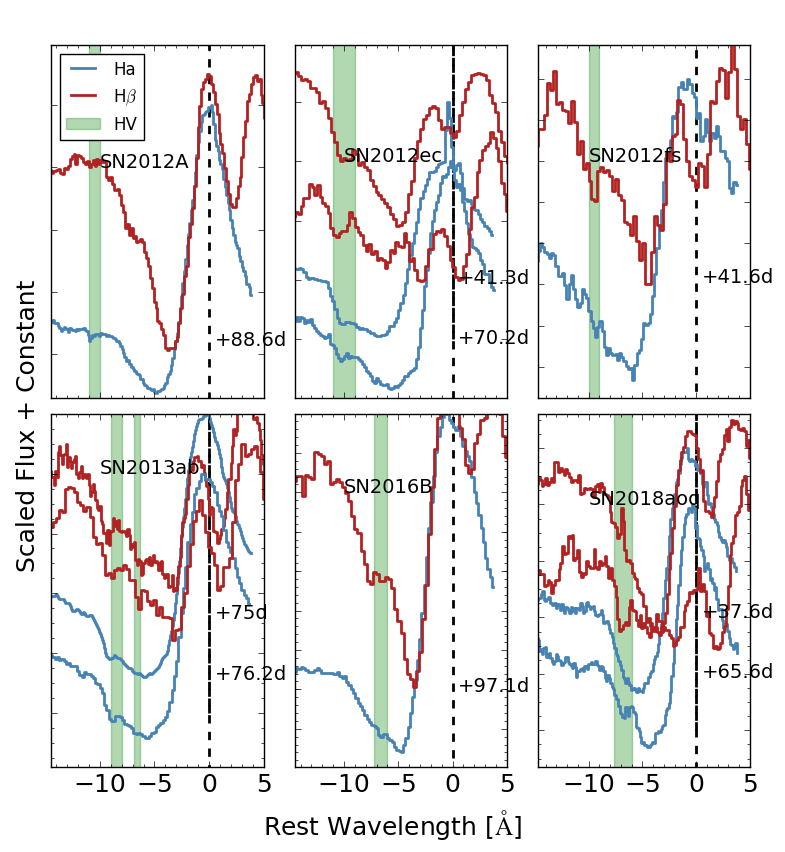}
    \caption{High Velocity features of hydrogen identified in our sample. Blue lines represent line profile of H$\alpha$ while red lines represent that of H$\beta$. The high velocity notches are marked by green shadow region. HV features have different shape: the notch of SN 2012A and SN 2016B are shallower, while SN 2012ec, SN 2012fs, SN 2018aoq are deeper. We found two notches in the $t\sim76.2$d spectrum of SN 2013ab at 76.2d, with the absorption minima locating at $9000\ {\rm km\ s^{-1}}$ and ${\rm 6500\ km\ s^{-1}}$, respectively. The overplotted $t\sim 75$d spectrum of SN 2013ab is taken from \citet{2012MNRAS.425.1789S}.} 
    \label{fig:HVspectra_2}
\end{figure}

For our sample presented in this work, we examined the blue part of H$\alpha$ in each spectrum and consider the notch as HV feature of H$\alpha$ if there is a similar absorption in ${\rm H\beta}$ line profile. The spectra with clear HV features are shown in Figure \ref{fig:HVspectra_2}, where one can also see that these HV features show significant diversities. 
The notches seen in SN 2012A and SN 2016B are shallower, while those seen in SN 2012ec, SN 2012fs and SN 2018aoq are deeper. 
As CSM interaction can produce different shapes of HV features, we suggested that CSM interaction be more likely to produce the HV features.

We noticed that SN 2013ab likely showed two HV features in $t\sim 76$ day spectrum, with the absorption minma locating at 
${\rm 9000\ km\ s^{-1}}$ and ${\rm 6500\ km\ s^{-1}}$, respectively. We further examed its $t\sim 75$ day spectrum from \citet{2012MNRAS.425.1789S} and found that the H$\beta$ did have two distinct absorptions, though the corresponding absorption in H$\alpha$ is very weak at ${\rm 6500\ km\ s^{-1}}$ and we were not sure if this absorption really existed.
Nevertheless, multiple notches are found to exist in other individuals. 
For instance, SN 2001X seems to have two notches detected in the same spectra (see Figure 18 in \citealt{2014MNRAS.442..844F}). However, \citet{2014MNRAS.442..844F} noticed that such an identification of HV feature is dubious as the velocities of H$\alpha$ and H$\beta$ do not match. 
Another sample is SN 2009N, where two absorption features next to H$\alpha$ (i.e., an absorption with a velocity of about $\rm 8000\ km\ s^{-1}$ and another absorption with a lower velocity) were detected at the same time in its spectra ranging from 62 days to 77 days after the explosion\citep{2014MNRAS.438..368T} (see their Fig 9). 
In SN 2009ip, four notches are found to embeded in the blue side of H$\alpha$ and its polarization spectra reveal a possibly clumpy line forming region.
It is possible that some notches shown in the spectra of these objects belongs to other species instead of hydrogen. Othewise, multiple HV features may indicates complex  structure of line forming region which is possibly produced by interactions with asymmetric CSM.

\subsection{Flashed-ionization Features}
\begin{figure*}
	\includegraphics[width=1.9\columnwidth]{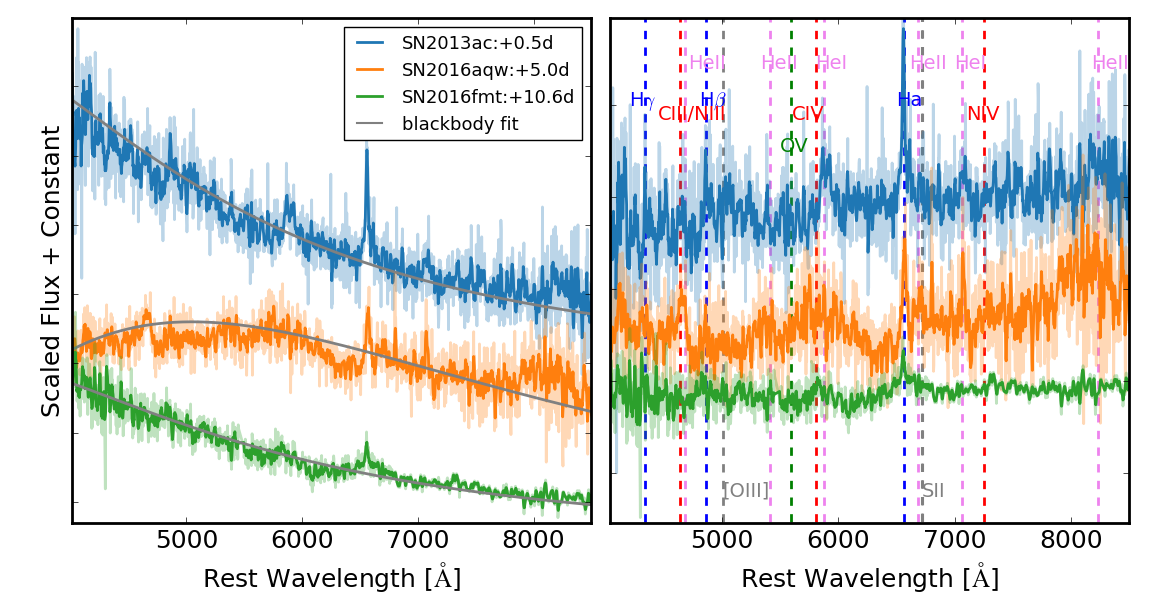}
    \caption{The SNe II of our sample showing possible flash ionized features in the early phases. Left panel: the original spectra and the best-fit blackbody (gray line). Right panel: the blackbody continuum-subtracted spectral features. Vertical dashed lines represent emission lines of different species.}
    \label{fig:FI_bbfit}
\end{figure*}

\citet{2016ApJ...818....3K} estimated that $\geq 18\%$ of SNe II tend to show ionization emission lines (flash ionized, FI) in their spectra if they are observed at sufficiently early times. Such FI features are formed by the recombination of the outermost CSM, which has been ionized by high-energy photons created during SN shock breakout \citep{2014Natur.509..471G}. 
In our sample, narrow H$\alpha$ emission is primarily used to identify the flash-ionized signatures as it is stronger and lasts for longer time. Ionization emission lines of other species, including helium, carbon, nitrogen and oxygen are also used whenever possible. 
As narrow emission lines of hydrogen can be easily contaminated by host-galaxy H II region, the objects with clear emission lines of host galaxy (e.g. [O~{\sc iii}] $5007$, S~{\sc ii} $6717, 6730$) but without emission lines of other species except hydrogen, are not assigned as SN II sample with FI features (e.g. SN 2016hvu, see Figure \ref{fig:sn_spec}). 
Finally, we identified possible FI features in two objects, i.e., SN 2013ac and SN 2016aqw, among the 49 objects with spectra taken within 11 days after the explosion, (see Figure \ref{fig:FI_bbfit}). The narrow emission lines of He~{\sc i} $5876,7065$ were observed in SN 2013ac and SN 2016aqw. 
In addition, the He~{\sc ii} $4686$ line (may blend with C~{\sc iii}/N~{\sc iii}) can be observed in the spectra of SN 2016aqw. Lorentzian function is used to measure the full width at half maximum (FWHM) of H$\alpha$ line profile, with the FWHM being $\sim 1000\ \rm km \ s^{-1}$ and $\sim \rm 800\  km \ s^{-1}$ for SN 2013ac and SN 2016aqw, respectively (see Figure \ref{fig:emi_plot}).
Note that for SN 2013fs, the FI feature can not be detected in our 5.4d spectrum, however, the emission features of O~{\sc vi} $3811$, O~{\sc v} $5597$, O~{\sc iv} $3410$, N~{\sc v} $4604$ and He~{\sc ii} $4686$ were detected in spectra that taken at earlier epochs than we observed \citep{2017NatPh..13..510Y}. 
Broad emission (with FWHM $\sim 5000\  \rm km\ s^{-1}$) has been detected at the location of H$\alpha$ in SN 2016fmt. However, a narrow component seems to be superposed on the broad component (see Figure \ref{fig:emi_plot}) and a possible emission signature of C~{\sc iv} 5801 seems to be detected, suggesting that SN 2016aqw possibly has FI features.

Besides the object with FI features, 19 of all 49 objects only show blue featureless spectra at early times ($\leq 11$\,d). The phases of these spectra range from 1.1 days to 9.3 days after explosion. As the FI features usually disappear quickly, detection of such features require observations of SNe II within a few days after their explosions. SNe II with only blue featureless continuum probably
produce no FI features at all or show FI signatures in the spectra at an epoch earlier than we observed. 

\section{Expansion Velocity}
\begin{figure*}
	\includegraphics[width=1.9\columnwidth]{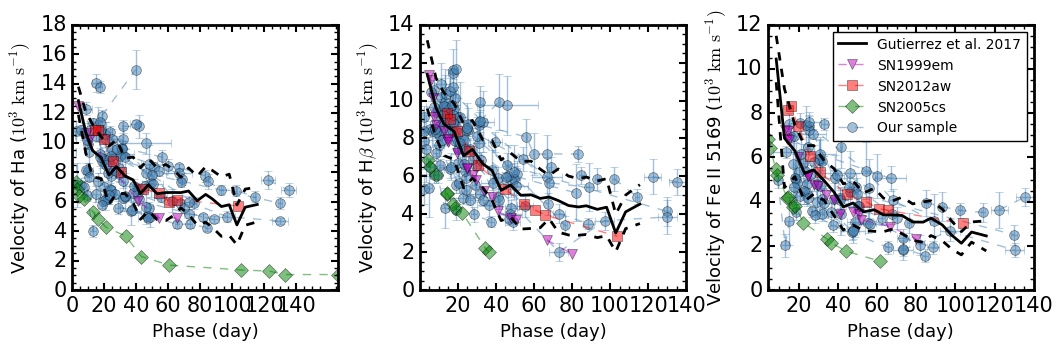}
    \caption{Velocity evolution of ${\rm H\alpha}$, ${\rm H\beta}$ and  Fe~{\sc ii} 5169 lines of our SN II sample and some well-studied SNe II. The blue circles represent our sample, while the black solid line and dashed lines represent the average velocity evolution and $1-\sigma $ standard deviation from the sample of \citet{2017ApJ...850...89G}.}
    \label{fig:velocityplot_hahbfe_all}
\end{figure*}

We measured the velocities of ${\rm H\alpha}$, ${\rm H\beta}$ and Fe~{\sc ii} 5169 from the absorption minima of P Cgyni profiles in every spectrum where they can be clearly detected.
At around the absorption minima, we choose a few regions with lower spectral flux density to measure the velocity. The average value of these measurements is used as the velocity and the standard derivation is used as the uncertainty.
The results are shown in Figure \ref{fig:velocityplot_hahbfe_all} and tabulated in Table \ref{table:spec_info}. For comparison, we also included the average velocity evolution from the sample of \citet{2017ApJ...850...89G}, the typical IIP SN 1999em \citep{2003MNRAS.338..939E,2012MNRAS.419.2783T}, the high-luminosity type IIP SN 2012aw \citep{2013MNRAS.433.1871B} and the low-luminosity SN~IIP 2005cs (LLSNe) \citep{2009MNRAS.394.2266P}. 
In Figure \ref{fig:velocityplot_hahbfe_all}, we see that all of our SNe except for SN 2016cok have velocities larger than SN 2005cs, which means that SN 2016cok could be a low-luminosity SN II. 
\citet{2017MNRAS.467.3347K} identified a pre-explosion counterpart to SN 2016cok in archival images of Hubble Space Telescope and they found that the initial mass of progenitor was most likely in the mass range of 8 $\sim$ 12 ${M_\odot}$. Such a relative lower mass range is consistent with that expected for the progenitor stars of low-luminosity SNe. 

\begin{figure*}
	\includegraphics[width=1.8\columnwidth]{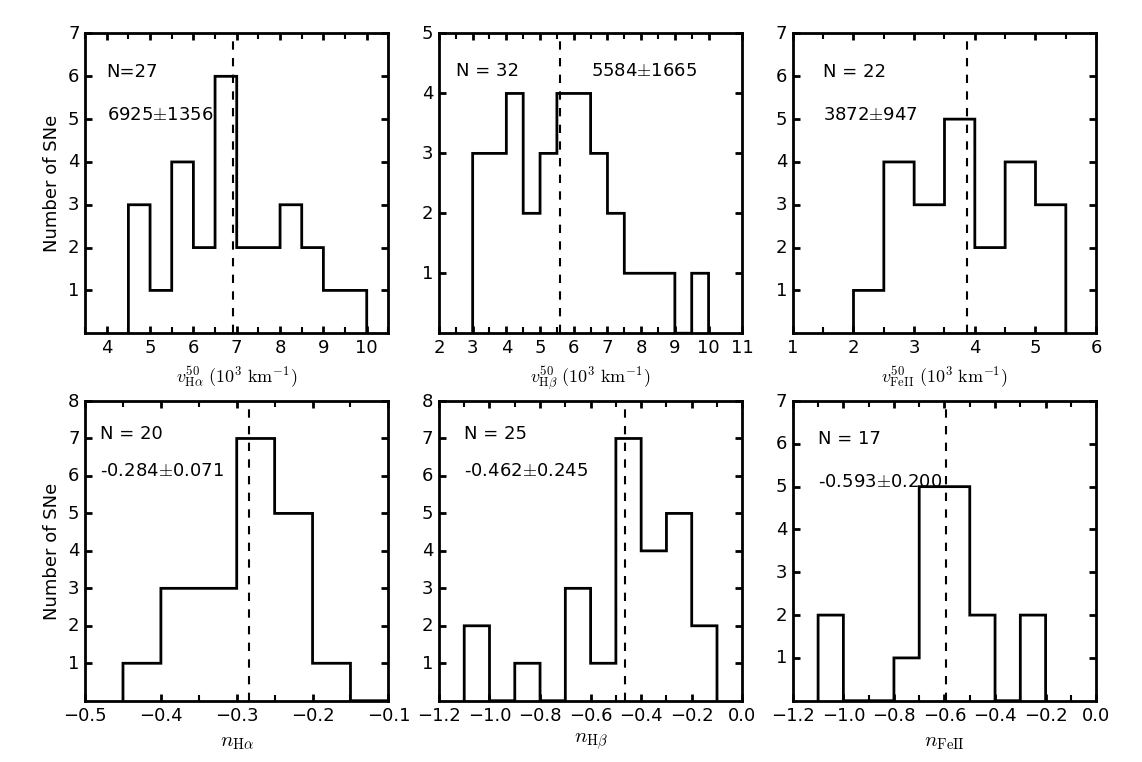}
    \caption{Distribution of velocity measured at $t\sim $50 days after the explosion and the power-law exponents of ${\rm H\alpha}$, ${\rm H\beta}$ and  Fe~{\sc ii} $5169$. The vertical dashed lines mark the average values. The total number and the average velocity (the unit of average velocity is $\rm km\ s^{-1}$) of each parameter are listed in each panel.}
    \label{fig:velocity_analysis}
\end{figure*}

As the SN ejecta expand homologously, thus the expansion velocities decrease with time approximately in a power-law fashion \citep{2014MNRAS.442..844F,2019MNRAS.490.2799D}.
We thus fit the velocity evolution with a power-law function (i.e., $v_{\rm {H\alpha}} = v^{50}_{\rm H\alpha}(t/{50})^{n_{\rm H\alpha}}$, $v_{\rm {H\beta}} = v^{50}_{\rm H\beta}(t/{50})^{n_{\rm H\beta}}$, $v_{\rm {Fe}} = v^{50}_{\rm Fe}(t/{50})^{n_{\rm Fe}}$) to estimate both the velocity at $t=50$ days after the explosion and the power-law exponent. 
Power-law fitting can only be used for those having at least two velocity data points. In our sample, however, many SNe have only one spectrum. If their spectra were taken between 45 to 55 days after the explosion, the velocity measured by absorption minimum can be treated as v50 and they are included in our analysis to increase the number of statistics.
The  parameters of $v_{50}$ and the power-law exponent are shown in Table \ref{tab:v50_n}.
The velocity evolution of a part of SNe II and the fitting results are shown in Figure \ref{fig:velocity_show_fit}, and the distribution of these parameters are shown in Figure \ref{fig:velocity_analysis}. 
At $t \sim 50$\,days after the explosion, the velocities have large varieties. For instance, the velocity of H$\alpha$ ranges from 4500 to 10,000 ${\rm km\ s^{-1}}$, with a mean value of $6904 \pm 1336\ {\rm km\ s^{-1}}$, while the velocities of  H$\beta$ range from 3000 to 10,000 ${\rm km\ s^{-1}}$ and the velocity of Fe~{\sc ii} 5169 range from 2000 to 5500 ${\rm km\ s^{-1}}$. 
As seen from Figure \ref{fig:velocity_show_fit}, for each SN, the velocity of hydrogen is higher than that of iron at similar epoch. This can be explained as that the Balmer lines are formed at larger radii than the iron lines. 

\citet{2017ApJ...850...89G} introduced $\Delta v({\rm H\beta})$ (the mean velocity decline rate in a fixed phase range) to quantify the decline rate of H${\beta}$ vecloity. They found that SNe II with larger decline rates at early times continue to show such behaviors at late times.
In our analysis, we use the power-law exponent (e.g., $\rm n_{H\alpha}$, see Section 4) to describe the velocity decline rate. Larger absolute values suggest faster decline rates. 
As seen from Figure \ref{fig:velocity_analysis}, the exponent measured from ${\rm H\alpha}$ velocity evolution ranges from -0.45 to -0.15 and has a smaller range compared with those inferred from H$\beta$ and Fe~{\sc ii} 5169 velocities. For example, the exponent inferred from the H$\beta$ velocity evolution has a range of -1.1 $\sim$ -0.15.
Among our sample, SN 2018bek, SN 2013gd and SN 2011az are the three showing the fastest decline rate of H$\beta$ velocity. 
The exponent derived from Fe~{\sc ii} 5169 velocity is found to vary from -1.1 to -0.2, with SN 2015V and SN 2016jfu are the two showing the fastest decline rates and SN 2011bi and PS15cwo showing the slowest decline rates. 
However, it should be pointed out that the power-law exponent was derived from only a few data points, which may suffer large uncertainties. Sometimes we may overestimate or underestimate the decline rate. For example, the $\rm n_{H\beta}$ of SN 2013gd,  $\rm n_{Fe}$ of SN 2015V and  $\rm n_{Fe}$ of SN 2016jfu may be overestimated as shown in Figure \ref{fig:velocity_show_fit}.


\begin{figure*}
	\includegraphics[width=1.8\columnwidth]{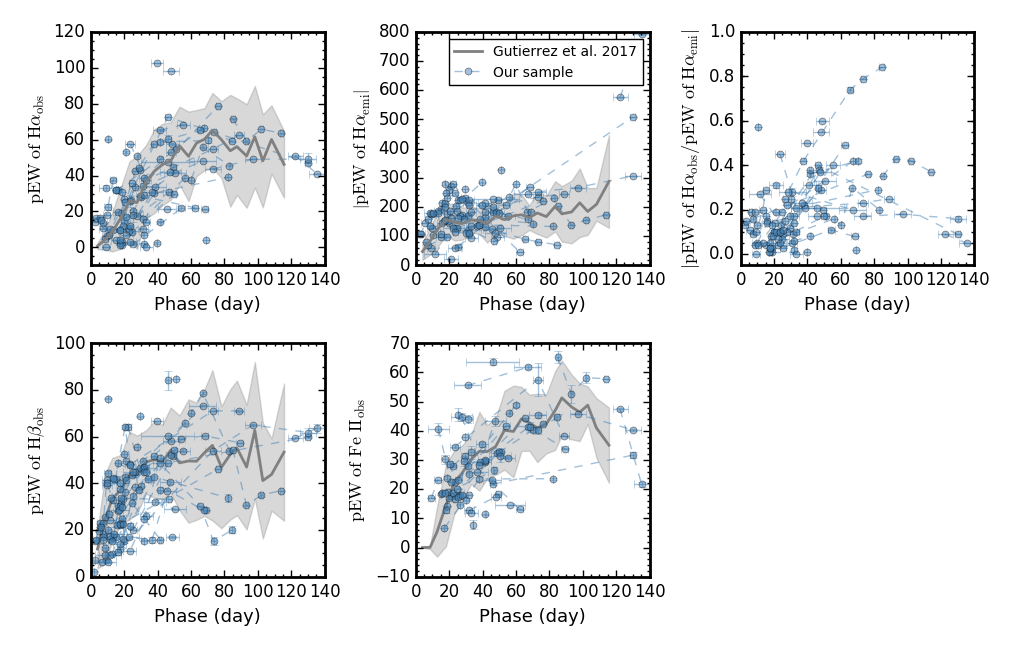}
    \caption{Temporal evolution of pEW of ${\rm H\alpha}$, ${\rm H\beta}$ and Fe~{\sc ii} 5169. The blue dots and the gray shadow represent the average pEW value and $1\sigma$ standard deviation from the sample given by \citet{2017ApJ...850...89G}}
    \label{fig:pew_plot_all_hahbfe}
\end{figure*}

\begin{figure*}
	\includegraphics[width=1.9\columnwidth]{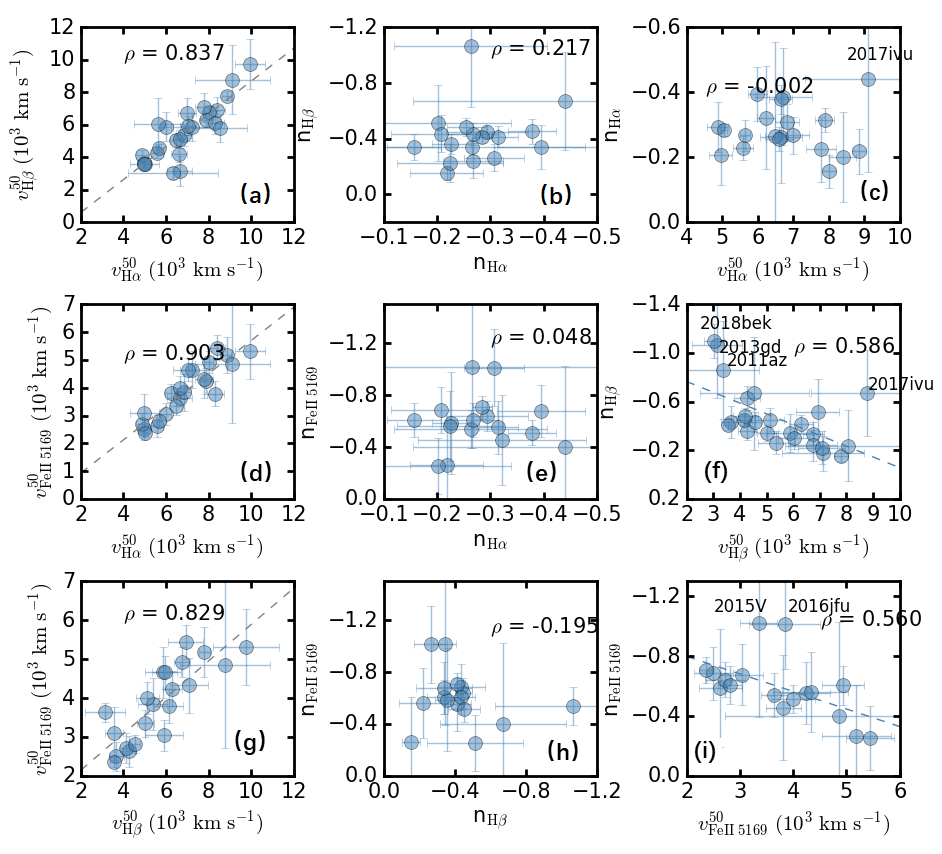}
    \caption{Correlations between velocity evolution parameters inferred from H$\alpha$, H$\beta$, and Fe~{\sc ii} 5169 lines. $v^{50}$ represents the velocity derived from $t\sim$ 50d spectra, while $n_{\rm H\alpha}$, $n_{\rm H\beta}$ and $n_{\rm Fe}$ represents the power-law index determined by applying the power-law fits to the velocity decay from the H$\alpha$, H$\beta$ and Fe~{\sc ii} 5169 absortpion features, respectively. 
    } 
    
    \label{fig:cor1_cor2}
\end{figure*}

\begin{figure*}
	\includegraphics[width=1.9\columnwidth]{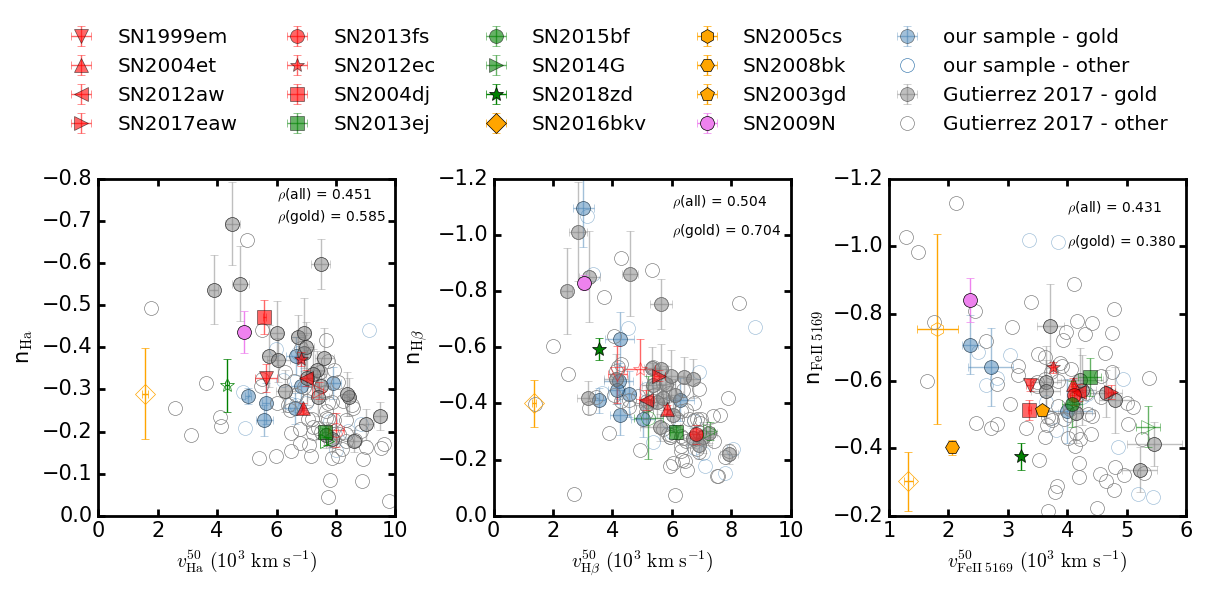}
    \caption{Similar as Fig \ref{fig:cor1_cor2}, with well-studied individuals in literature included. 
    Blue circles represent the data presented in this paper, while other well-studied sample are denoted by different symbols, with red symbols representing "typical" SNe IIP, green symbols representing fast-declining objects and yellow ones representing low-luminosity SNe II, respectively.The solid markers represent the gold sample which satisfies the criteria that the uncertainty is less than 20\%, while the open markers represent other objects.}.
    \label{fig:moreobject}
\end{figure*}

\begin{figure*}
	\includegraphics[width=1.2\columnwidth]{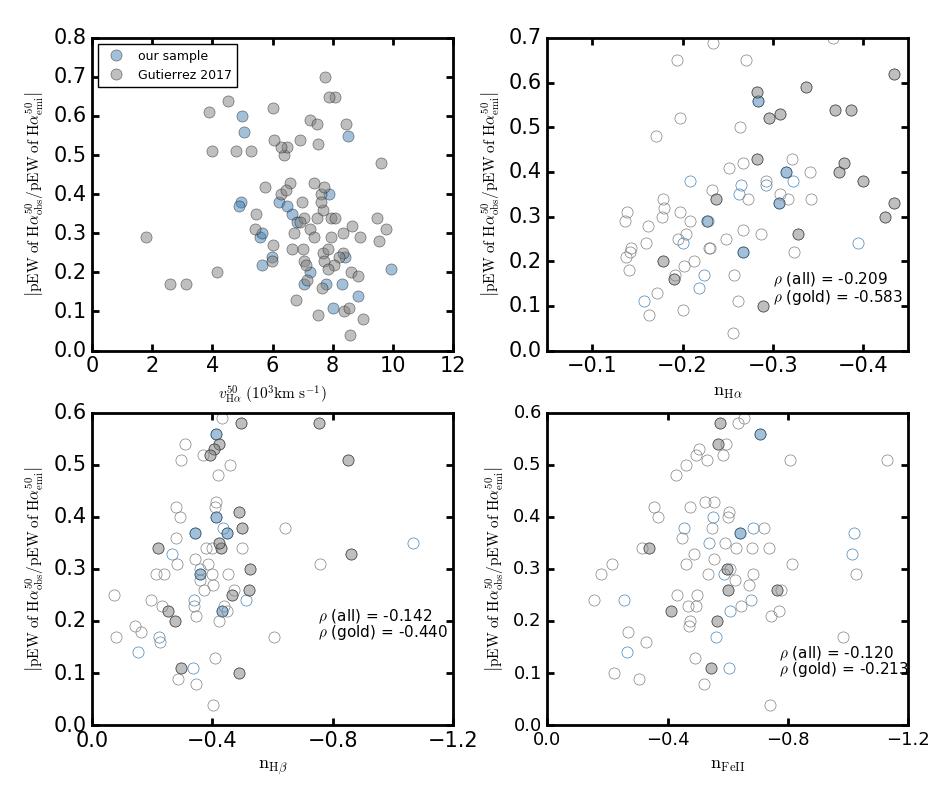}
	\caption{Correlation between pEW ratios of absorption to emission component (a/e) of H$\alpha$ and the velocity power-law exponents. In the upper-right, lower-left, and lower-right panels, the gold sample (The definition of a gold sample is the same as Figure \ref{fig:moreobject}) are represented by solid markers, while other objects are represented by open markers. }
	\label{fig:pew_n}
\end{figure*}

\section{Pseudo-equivalent Widths}
The pseudo equivalent width (pEW) is used to describe the strength of spectral lines. We measured the pEW of the absorption part of H$\alpha$, ${\rm H\beta}$ and Fe~{\sc ii} 5169 lines, the emission part of H$\alpha$ and the pEW ratio of absorption to emission component (a/e) of H$\alpha$. The measurement results of our sample and the average values from sample presented in \citet{2017ApJ...850...89G} are shown in Figure \ref{fig:pew_plot_all_hahbfe}. 
One can see that the measurements of our sample are consistant with those derived by \citet{2017ApJ...850...89G}, although it is complicated to determine by the underlying continuum of each line feature.\footnote{In our work, the continuum is determined by a linear fit to the two ends of the spectral line and the uncertainty of pEW is estimated by changing the range of two ends.} 
The pEW of absorption component can reach at $\sim 100$ \AA \ , $\sim 80$ \AA \ , and $\sim 70$ \AA \ for H$\alpha$, H$\beta$ and Fe~{\sc ii} 5169, respectively, while the emission component of H$\alpha$ can reach at 
$\sim 300$ \AA \ in the first $\sim$140 days.
As noted by \citet{2019MNRAS.490.2799D} and \citet{2017ApJ...850...89G}, generally speaking, 
the absorption components of Balmer lines and Fe II lines tend to increase their strengths with time in the first one or two months and this increase behavior becomes slowly at later phases. 
However, the pEW evolution shows diversities among different objects. 
For example, the absorption part of H$\alpha$ in PS15cwo, SN 2013gd and SN 2018aoq start to decrease in strength at $t \sim $ 60-70 days after explosion, while this happened in SN 2016cok since $t\sim 15$ days after the explosion. 
Diversities are also observed in the evolution of H$\alpha$ emission component. The evolution shows an overall increase with time but several objects show a decrease trend at some point.
The general tendency of a/e shows an increasing tendency at early times and it remains constant or shows slight decrease at late times. 
The large diversities in the strength of the above lines among different SNe II may reflect diverse temperature evolution and progenitor 
metallicity \citep{2017ApJ...850...89G}. 


\section{PARAMETERS of Velocity Evolution}

\subsection{Correlations between velocity parameters}
For these velocity parameters measured for our sample, we use the Pearson correlation coefficient ($\rho$) to examine their correlations. 
According to \cite{Evans1996}, the coefficients in the range of 0-0.19, 0.2-0.39, 0.4-0.59, 0.6-0.89 and 0.8-1.0 represent no, weak, moderate, strong, and very strong correlations, respectively. 
The results are shown in Figure \ref{fig:cor1_cor2}. 
For the velocities measured at 50 days after the explosion, we found that any two of $v^{50}_{\rm H\alpha}$, $v^{50}_{\rm H\beta}$ and $v^{50}_{\rm FeII 5169}$ parameters show a strong positive correlation, with the Pearson correlation coefficient being larger than 0.8 (see the left three panels of Figure \ref{fig:cor1_cor2}). However, for the evolution exponent $n_{\rm Ha}$, $n_{\rm H\beta}$ and $\rm n_{Fe}$, we find no correlation between them, as shown in the middle panels of Figure \ref{fig:cor1_cor2}.
Besides, we found that the ${\rm H\beta}$ velocity tends to show a slower decline for SNe II with higher velocities. (see panel (f) of Figure \ref{fig:cor1_cor2}). 
After excluding SN 2018bek, SN 2013gd and SN 2011az, which are the three showing the fastest velocity declines\footnote{The point of SN 2017ivu is also removed since it is an oulier which probably due to bad spectral sampling.}, the Pearson coefficient increase froms 0.586 to 0.660.

To test whether the above negative tendency really exists for SNe II, we further include some well-studied individual SNe II from the literature and those of \cite{2017ApJ...850...89G} in the statistical sample. The velocity evolution and power-law fit for these well studied objects are shown in Figure \ref{fig:velocityshow_compareobject} and the relevant parameters of these well-studied sample are also given in Table \ref{tab:v50_n}. The power-law fitting and the parameters for the sample of \citet{2017ApJ...850...89G} are shown in Figure \ref{fig:gutierrez_powerlaw_fit} and Table \ref{table:gutierrez_v50_n}, respectively.
For each spectral species, Pearson correlation coefficients were calculated for all the samples and for only the gold sample, respectively. (Those with error of power-law exponent less than 20\% are selected as the gold sample and they are represented by solid markers in Figure \ref{fig:moreobject}.)
As can be seen from Figure \ref{fig:moreobject}, the tendency that high-velocity SNe II have slower velocity gradients still remains for ${\rm H\beta}$ when including more objects. 
Similar tendency probably exists in ${\rm H\alpha}$, with the Pearson coefficient being 0.45 for all sample and 0.65 for the well-observed SNe II from literature (after removing SN 2016bkv). 
\citet{2014MNRAS.445..554F} found that the velocities inferred from hydrogen lines of SNe IIL, especially ${\rm H\beta}$, evolve more slowly than those of SNe IIP. 
In general, the velocities of SNe IIL are higher than those of SNe IIP, which means that SNe II with higher H$\beta$ velocities also evolve more slowly in the samples of \citet{2014MNRAS.445..554F}. 
Therefore, what they found is consistent with the negative correlation in H$\beta$ revealed in this work. 
In comparison, the velocity decline rate of hydrogen derived in this work shows a continuous distribution 
and no distinction is found between the whole SNe II smaple.

Besides, a negative correlation possibly exists in Fe~{\sc ii} velocity and its velocity decay . However, we caution, that the negative correlation is not evident in the well-studied sample and the parameter distribution of Fe seems to have a larger scatter than that of hydrogen.
Moreover, both \citet{2014MNRAS.445..554F} and \citet{2015ApJ...815..121D} suggested that the Fe~{\sc ii} velocity of SNe IIL shows a similar evolution as that of SNe IIP (but with a significant scatter), which is not consistent with the negative correlation.
Therefore, we can not conclude the correlation between Fe~{\sc ii} velocity and its decay rate.

As shown in Figure \ref{fig:moreobject}, the velocity parameters of some objects deviate from others.
In some cases, the power-law results seem unreasonable due to bad spectral sampling (e.g., SN 2015V and SN 2016jfu).
In other cases, the velocities evolve in distinct ways. 
The velocities of both SN 2016bkv (a low-luminosity SN II) and SN 2018zd show an increase at early times, which may be related to CSM interaction\citep{2020MNRAS.498...84Z}, as shown in Figure\ref{fig:velocityshow_compareobject}. 
Moreover, the velocity parameters of SN 2016bkv seem to show significant differences from others. 
\citet{2018ApJ...859...78N} also noticed that SN 2016bkv showed a slower velocity evolution when compared with other LL SNe.
It is not clear why some LL SNe II deviate from the "normal" SN II sample in the $v^{50} - n$ relation.
The evolution of velocity is related
to the physical properties of the progenitor star and expansion itself. Since SN 2016bkv could be the possible electron-capture supernovae candidate \citep{2018ApJ...861...63H},its distinct velocity evolution may be partly related with the possible electron-capture origin.  

\subsection{Correlations between pEW and velocity}

The ratio of absorption part to emission part (a/e) of H$\alpha$ can be used to describe the diversity of SNe II. \citet{1994A&A...282..731P} and \citet{1996AJ....111.1660S} proposed that the subclass of SNe IIL have shallower P-Cgyni profiles, and hence smaller a/e values. 
Whereas \citet{2014ApJ...786L..15G} proposed that there is no definitive spectral distinction between SNe IIP and SNe IIL, and SNe II with smaller a/e ratios of H$\alpha$ tend to have higher H$\alpha$ velocities, more rapid post-peak declines in light curve, higher peak brightness, and shorter optically thick phase duration (defined as time from the explosion epoch through the end of the plateau phase.). 
Our spectral dataset also indicate that SNe II with higher H$\alpha$ velocities have smaller a/e ratios. 
Moreover, the exponents of velocity decay inferred from hydrogen lines seem to have a moderate correlation with the a/e values 
(see Fugure \ref{fig:pew_n}), suggesting that the velocity evolution of hydrogen lines may be related to the hydrogen envelope.
\cite{2014MNRAS.445..554F} proposed that the velocities of SNe IIL evolve more slowly than those of SNe IIP is due to that the hydrogen lines are formed in the outer thin layers of the ejecta rather than in a thick, gradually-exposed hydrogen envelope. The moderate correlation in hydrogen lines revealed by our sample (i.e., SN II with slower velocity decay of hydrogen lines tends to have a smaller a/e ratio) favours the above explanation and also favours the traditional consensus that those traditionally classified SNe IIL likely have smaller hydrogen envelope mass at SN explosion.

If SNe IIL are formed by single stars, less hydrogen envelope mass (larger mass loss) could point to higher progenitor mass than SNe IIP \citep{2009ARA&A..47...63S} or high progenitor metallicity at which mass loss is strong enough to remove much of the hydrogen envelope \citep{2003ApJ...591..288H}. 
Direct observations on progenitor stars of SNe IIL are rare and the estimated masses have large uncertainties. Nevertheless, \citet{2016MNRAS.459.3939V} proposed that SNe IIL appear not to come from more massive progenitors than SNe IIP. 
Moreover, single stars with enough metallicity will be so massive and they tend to form blackholes by fallback with weak supernovae \citep{2003ApJ...591..288H}.
If part of SNe IIL do not have larger progenitor mass than SNe IIP, then it is possible that they originate from binary system and the mass transfer in binaries may play important roles in stripping part of the hydrogen shell.
In future works, we intend to discuss the lightcurve properties and the correlations between spectral and photometric parameters, as well as the SN environment, aiming to understand the SN progenitor and its explosion physics.

\section{Conclusions}
In this work, we present a compilation of the optical spectra of SNe IIP/IIL observed over the past decade by the Tsinghua Supernova group. The sample consists 206 unreleased spectra for 104 SNe II, covering the phases from $\sim$1d to $\sim$200d after the explosion.  
Among 49 objects with spectra obtained within 10 days after explosion, two objects (SN 2013ac and SN 2016aqw) are found to show possble flash-ionized emission lines, including narrow lines of ${\rm H\alpha}$, ${\rm H\beta}, $HeI $5876,7065$, and  HeII $4686$. Prominent HV features of hydrogen are detected in six objects and they show different line profiles. Two are shallower and three are deeper. Two HV features are found to likely exist in the blue side of both H$\alpha$ and H$\beta$ in the $t\sim $ 76.2d spectrum of SN 2013ab. Based on the diversity of line profiles and multiple components, we suggested that CSM interactions could be at least a partial origin of HV features. 

The pEW of absorption component of H$\alpha$ in our sample can reach at $\sim$ 100 \AA \  and the emission component can reach at $\sim$ 300 \AA \ in the first  $\sim 140$ days.  
For individual object, the pEW of Balmer and metal lines show large scatter in the evolution, but we do not find definitive distinctions to separate SNe IIP from SNe IIL. 
A power-law function is used to fit the velocity evolution in order to estimate the velocity at 50 days after the explosion; moreover, the power-law exponent can be used to describe the velocity decay rate at the same time. 
For our sample, we found that SNe II with higher velocities during plateau phase show slower velocity evolution for H$\beta$. Moreover, the velocity decay rate of hydrogen (i.e., $\rm n_{H\alpha}$ and $\rm n_{H\beta}$) have a moderate correlation with the a/e of H$\alpha$. 
SNe II with smaller velocity gradient of hydrogen lines tend to have smaller a/e ratios as well, suggesting the progenitors of those SNe II having less amount of hydorgen kept in the stellar envelope before explosion. 
Photometric parameters, as well as SN environment, will be discussed in future works, in order to understand the observational diversity of SNe II and the mechanisms responsible for these diversities.

\section*{Acknowledgements}
We acknowledge the support of the staff of the Lijiang 2.4 m and Xinglong 2.16-m telescopes. This work is supported by the 
National Natural Science Foundation of China (NSFC grants 12288102, 12033002 and 11633002) and the Tencent Xplorer Prize. 
J. Zhang is supported by the National Key R\&D Program of China (2021YFA1600404), the National Natural Science Foundation of China (12173082),  the Yunnan Province Foundation (202201AT070069), the Top-notch Young Talents Program of Yunnan Province, the Light of West China Program provided by the Chinese Academy of Sciences, and the International Centre of Supernovae, Yunnan Key Laboratory (202302AN360001).
This work was supported by International Centre of Supernovae, Yunnan Key Laboratory (No. 202302AN360001).
Y.-Z. Cai is supported by the National Natural Science Foundation of China (NSFC, Grant No. 12303054).
This work is supported by the Strategic Priority Research Program of the Chinese Academy of Sciences, Grant No. XDB0550100.
This work is supported supported by the National Natural Science Foundation of China(Grant Nos. 12090041,12090040). 
This work was partially supported by the Open Project Program of the Key Laboratory of Optical Astronomy, National Astronomical Observatories, Chinese Academy of Sciences. 
The LJT is jointly operated and administrated by Yunnan Observatories and the Center for Astronomical MegaScience (CAS). Funding for the LJT has been provided by Chinese Academy of Sciences and the Peoples Government of Yunnan Province.


\section*{Data Availability}

The data underlying this article are available in the article and in its online supplementary material.
All the spectra have been uploaded to the webpage of WISeREP (\href{https://www.wiserep.org}{https://www.wiserep.org}) and Zenodo \href{https://doi.org/10.5281/zenodo.10466160}{(https://doi.org/10.5281/zenodo.10466160)}




\bibliographystyle{mnras}
\bibliography{example} 




\appendix

\section{Some extra material}


\begin{figure*}
	\includegraphics[width=2.2\columnwidth]{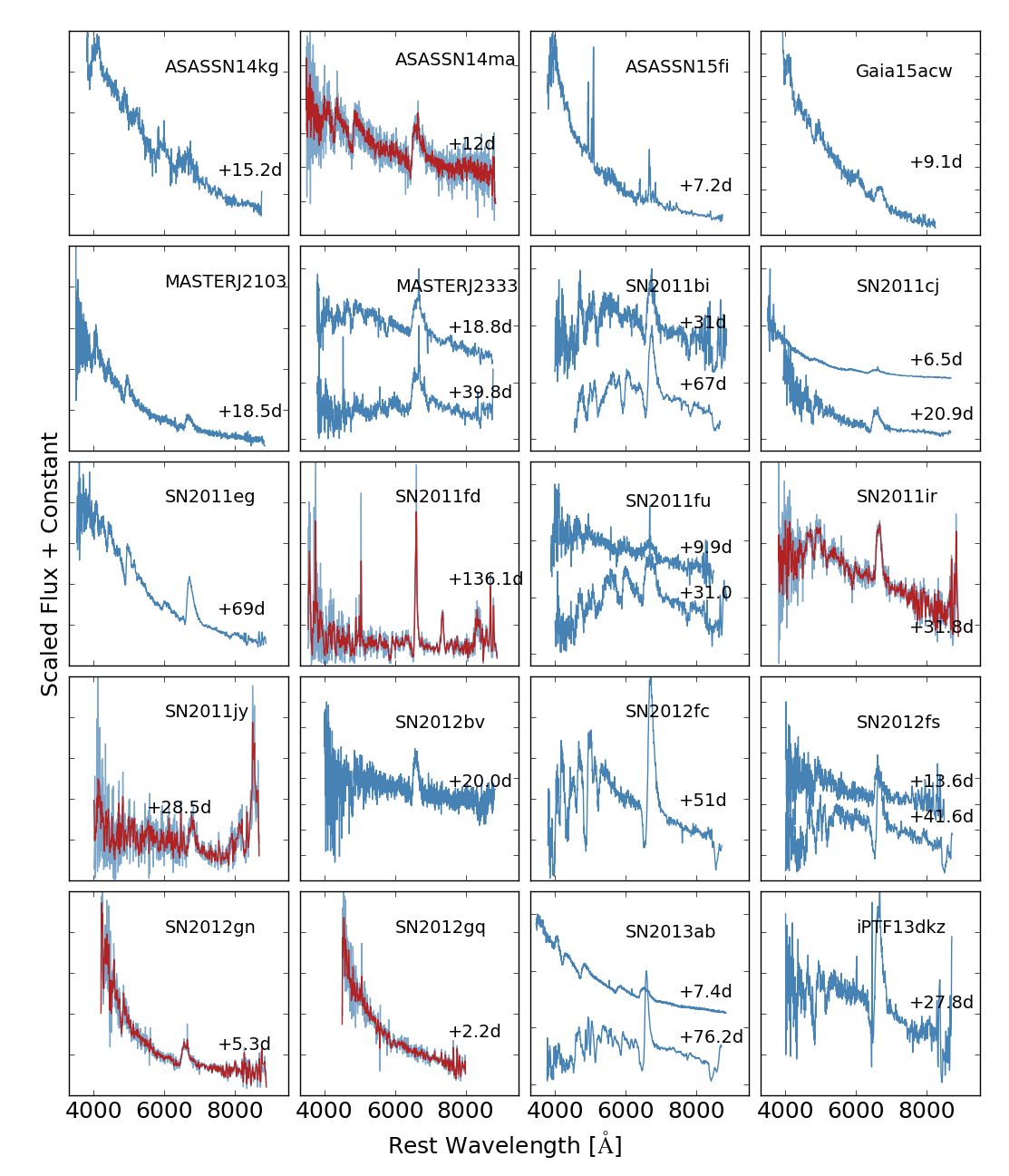}
    \label{fig:spec12_a1}
\end{figure*}

\addtocounter{figure}{-1} 

\begin{figure*}
	\includegraphics[width=2.2\columnwidth]{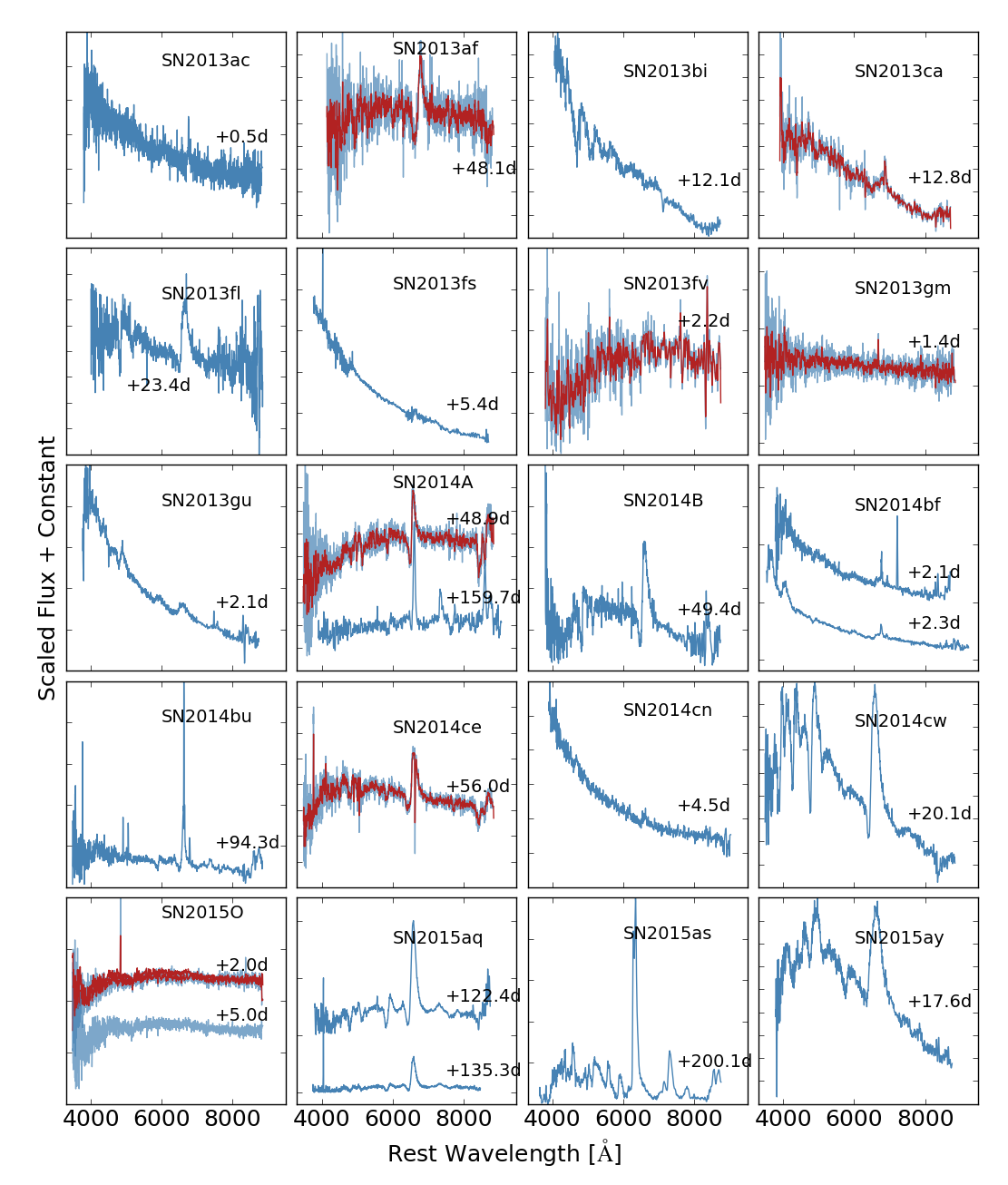}
    \label{fig:spec12_a1}
\end{figure*}

\addtocounter{figure}{-1} 

\begin{figure*}
	\includegraphics[width=2.2\columnwidth]{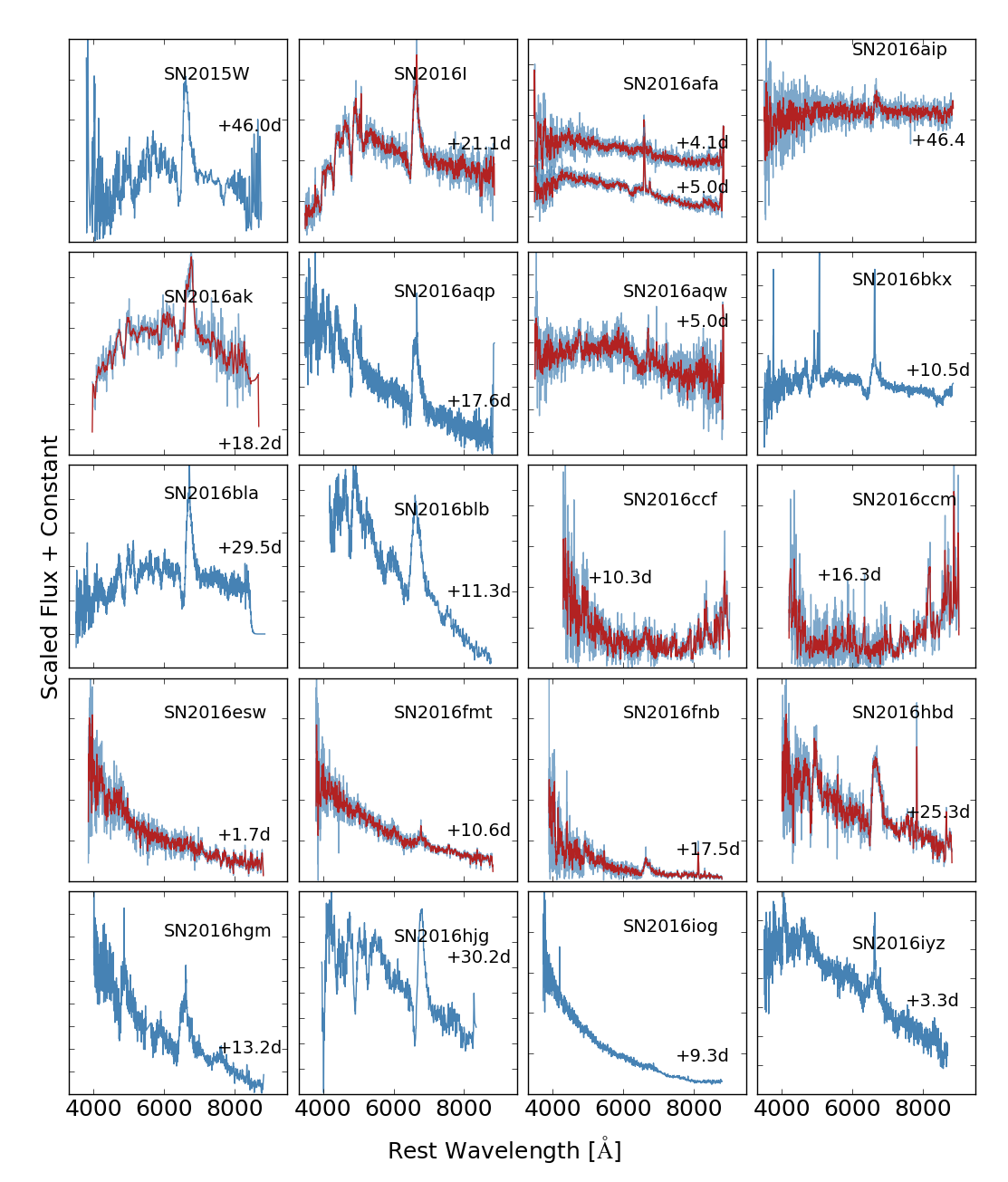}
    \label{fig:spec12_a1}
\end{figure*}

\addtocounter{figure}{-1} 

\begin{figure*}
	\includegraphics[width=2.2\columnwidth]{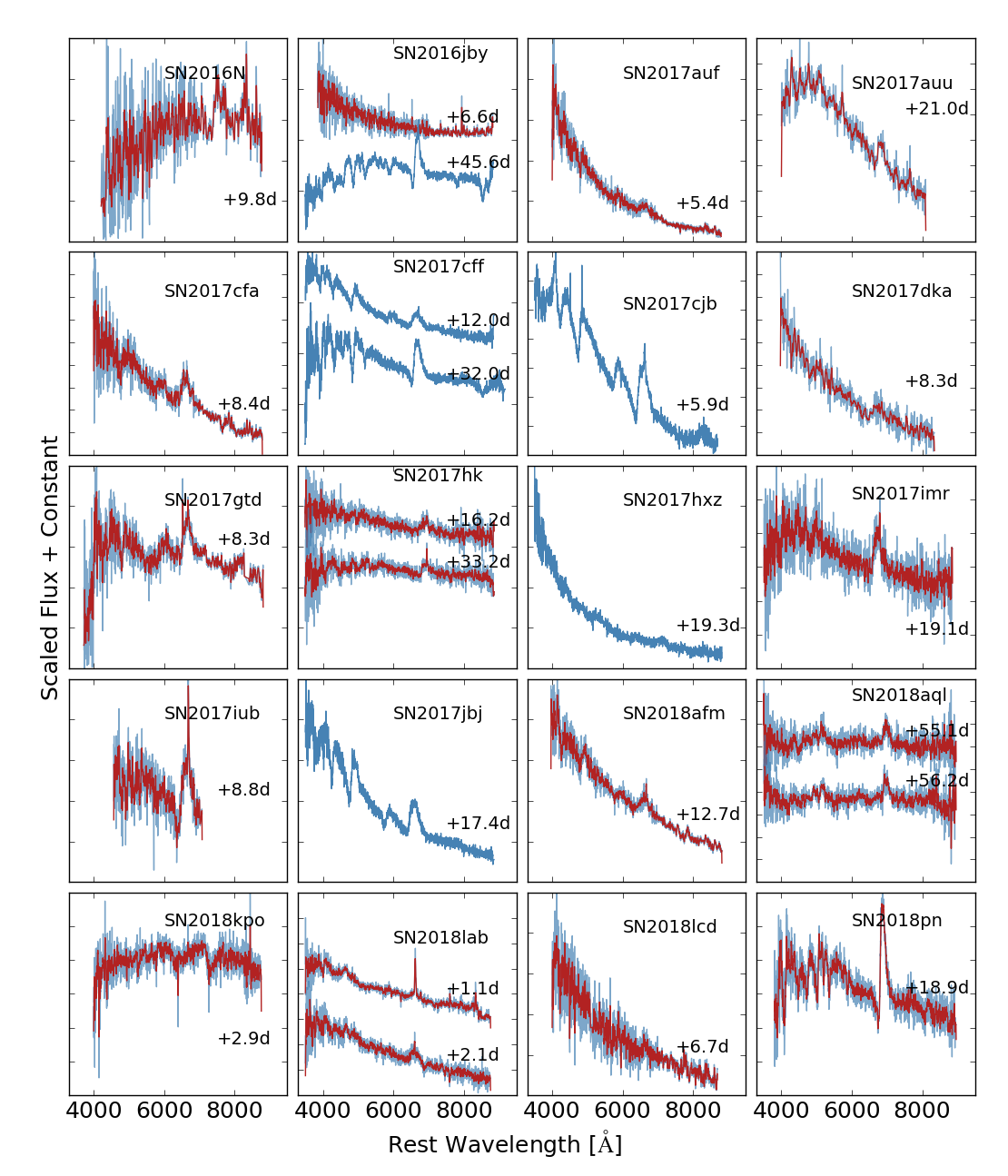}
    \label{fig:spec12_a1}
\end{figure*}

\addtocounter{figure}{-1} 

\begin{figure*}
	\includegraphics[width=2.2\columnwidth]{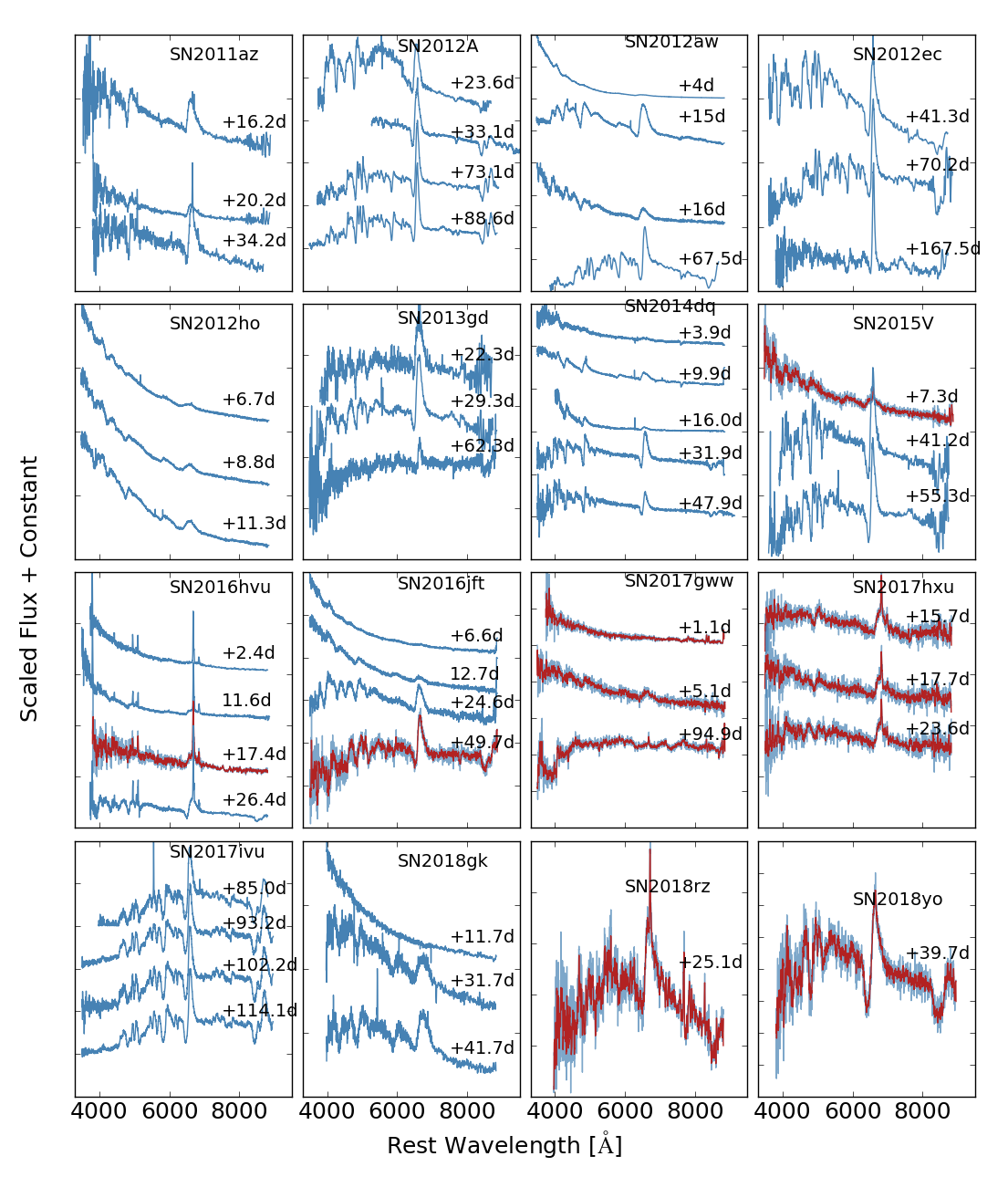}
    \label{fig:spec311_b1}
\end{figure*}

\addtocounter{figure}{-1} 

\begin{figure*}
    \addtocounter{figure}{3}
	\includegraphics[width=2.2\columnwidth]{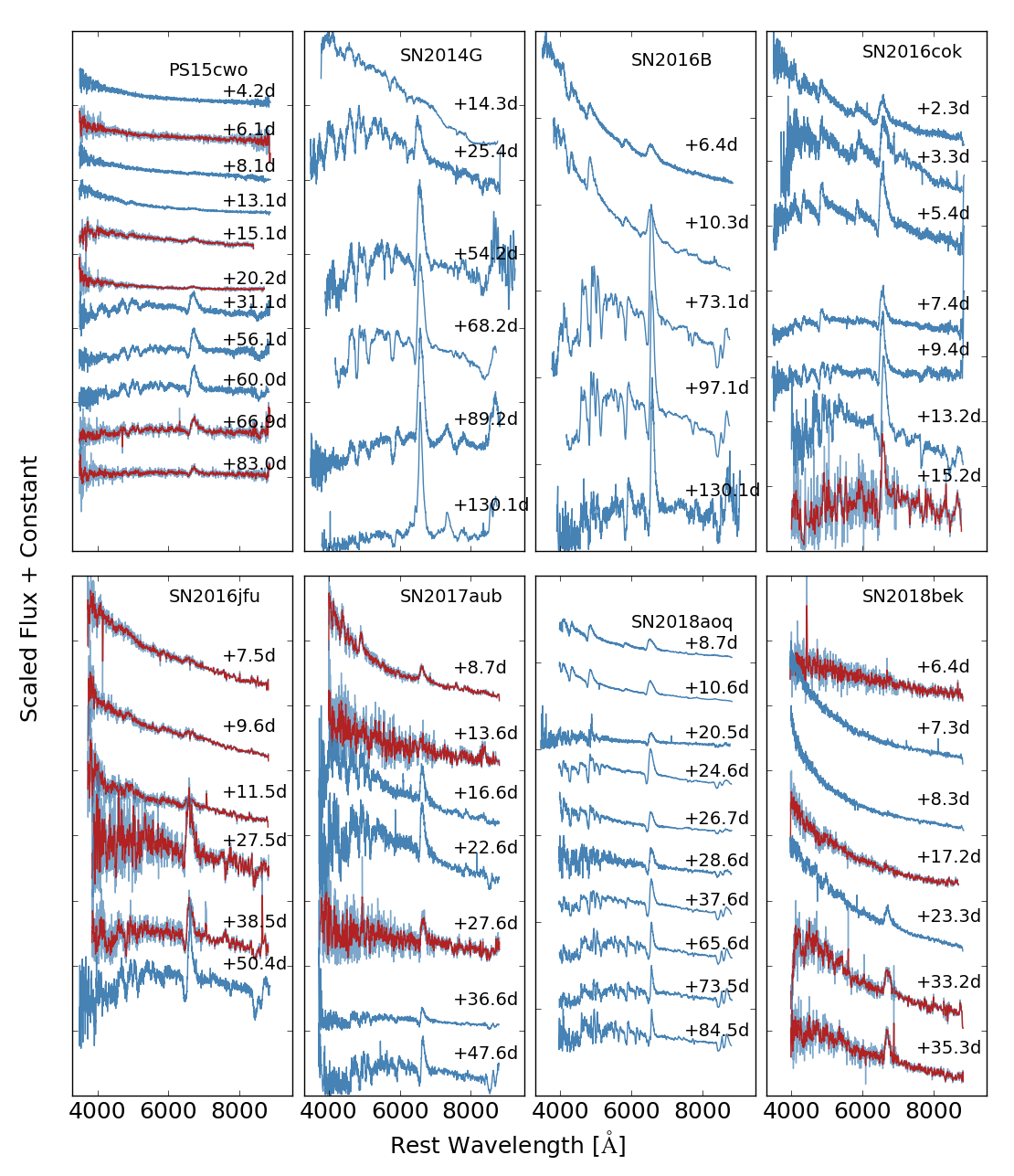}
    \caption{Spectral evolution of our SN II sample presented in this work.}
    \label{fig:sn_spec}
\end{figure*}

\begin{figure}
	\addtocounter{figure}{2}
	\includegraphics[width=0.9\columnwidth]{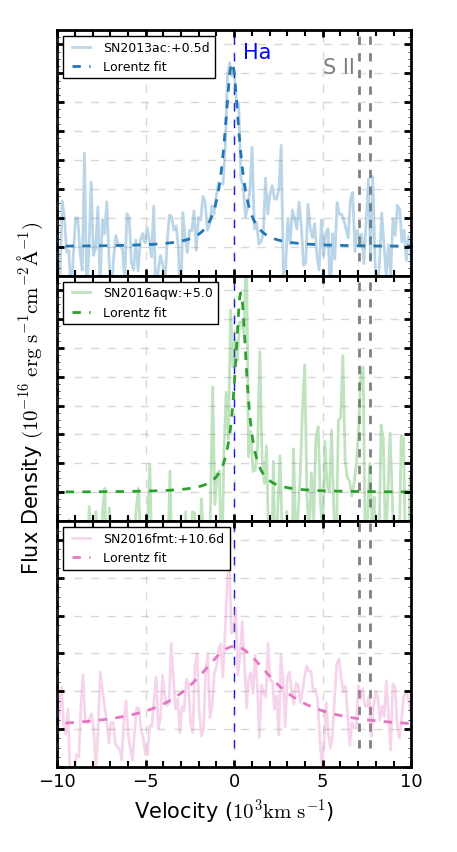}
	\caption{The H$\alpha$ line profile in the spectrum of SN2013ac, SN 2016aqw and SN 2016fmt. The dashed lines are the Lorentzian fits for the H$\alpha$ line profile.}
	\label{fig:emi_plot}
\end{figure}

\begin{figure*}
	\includegraphics[width=1.9\columnwidth]{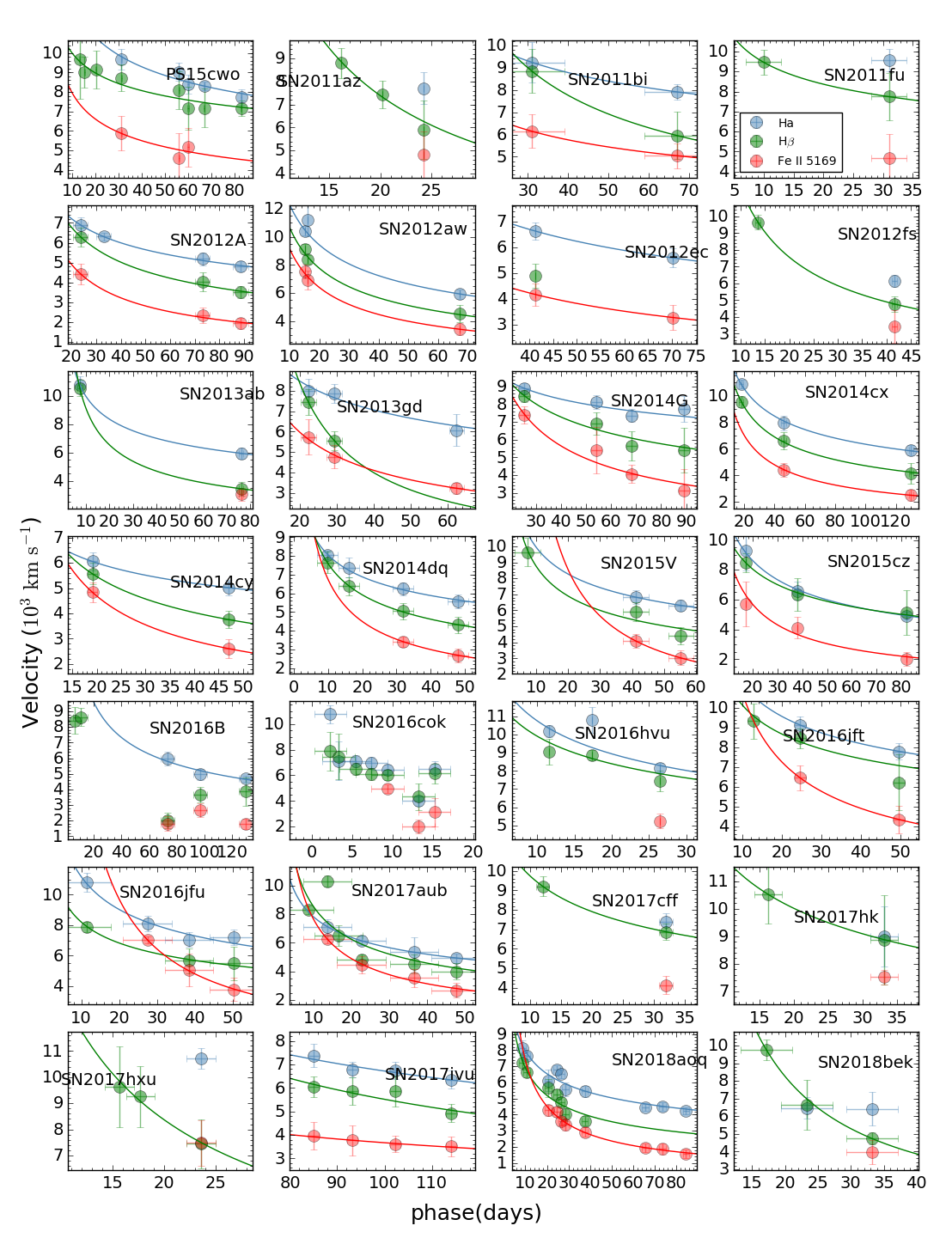}
    \caption{The velocity evolution inferred from ${\rm H\alpha}$ (blue dots), ${\rm H\beta}$ (green dots) and Fe~{\sc ii} 5169 (red dots) lines in the spectra of our sample of SNe II. Overplotted are the power-law fitting to the observed velocity evolution. 
    We must notice that the power-law exponent of our sample has larger errors, and this is due to bad spectral sampling. 
    }
    \label{fig:velocity_show_fit}
\end{figure*}

\begin{figure*}
	\includegraphics[width=1.9\columnwidth]{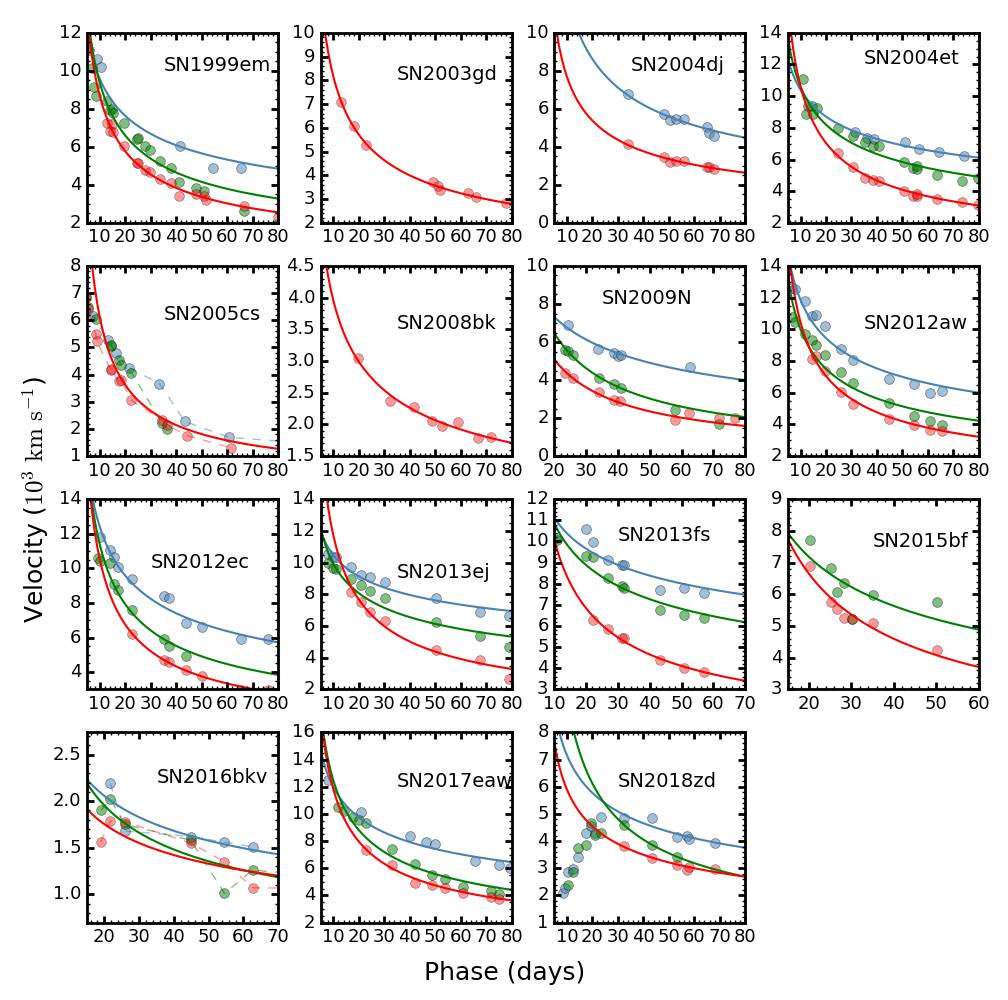}
    \caption{Same as Figure\ref{fig:velocity_show_fit}, but for the well-studied sample in literature. 
    The newly added sample include the typical SNe IIP like SN 1999em \citep{2003MNRAS.338..939E,2012MNRAS.419.2783T}, SN 2004dj \citep{2006MNRAS.369.1780V}, SN 2012ec \citep{2015MNRAS.448.2312B}, SN2004et \citep{2006MNRAS.372.1315S,2012MNRAS.419.2783T}, SN 2012aw \citep{2013MNRAS.433.1871B} and SN 2017eaw\citep{2019ApJ...875..136V,2019ApJ...876...19S}; 
    fast declining samples of SN 2013ej \citep{2015ApJ...807...59H} and SN 2015bf \citep{2021MNRAS.505.4890L});  
    the low-luminosity, low-velocity objects of SN 2003gd \citep{2005MNRAS.359..906H}, SN 2005cs \citep{2009MNRAS.394.2266P,2012MNRAS.419.2783T}, SN 2008bk\citep{2015IAUGA..2255784P} and SN2016bkv\citep{2018ApJ...859...78N};
    SN 2009N \citep{2014MNRAS.438..368T}, which link normal and subluminous SNe IIP; and SN 2018zd \citep{2020MNRAS.498...84Z}. 
    Among these objects, 
    SN 2013fs, SN 2015bf, SN 2016bkv and SN 2018zd have flash ionized features at early phases, which indicate massive CSM and therefore larger mass loss rate of progenitor shortly before the SN explosion.}
    \label{fig:velocityshow_compareobject}
\end{figure*}

\begin{figure*}
	\includegraphics[width=2.2\columnwidth]{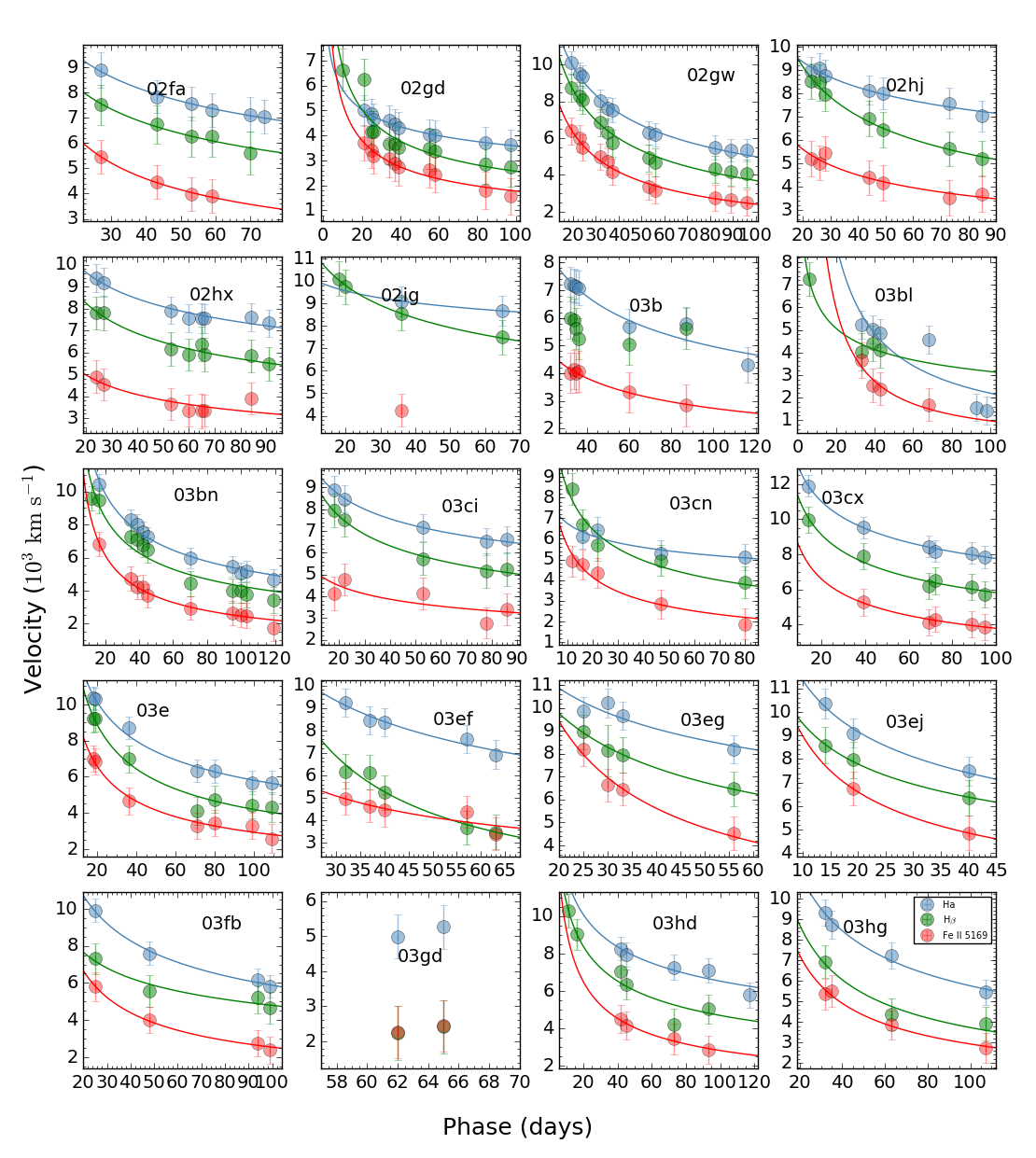}
	\label{fig:gplot_1}
\end{figure*}

\addtocounter{figure}{-1} 

\begin{figure*}
	\includegraphics[width=2.2\columnwidth]{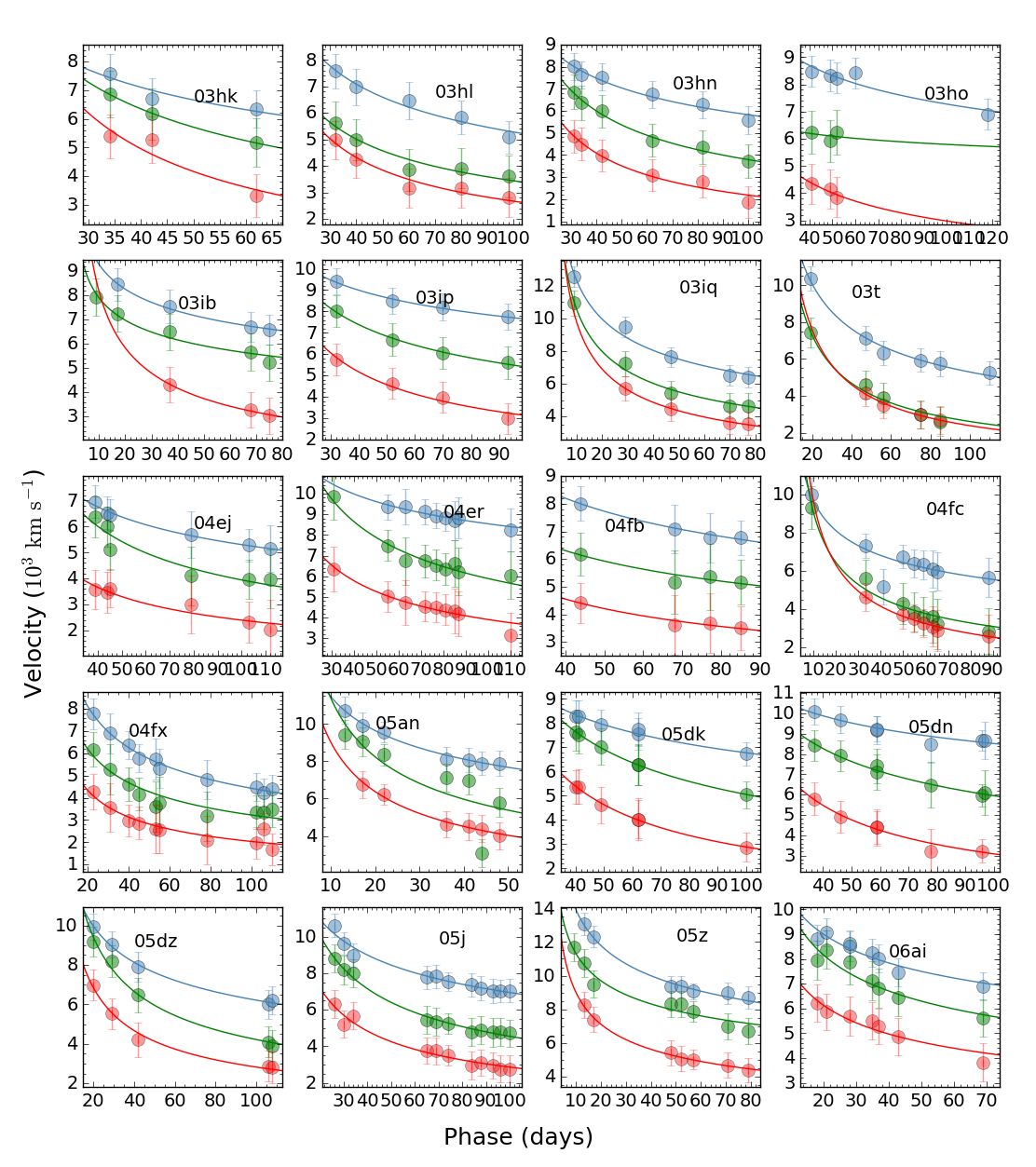}
	\label{fig:gplot_2}
\end{figure*}

\addtocounter{figure}{-1} 

\begin{figure*}
	\includegraphics[width=2.2\columnwidth]{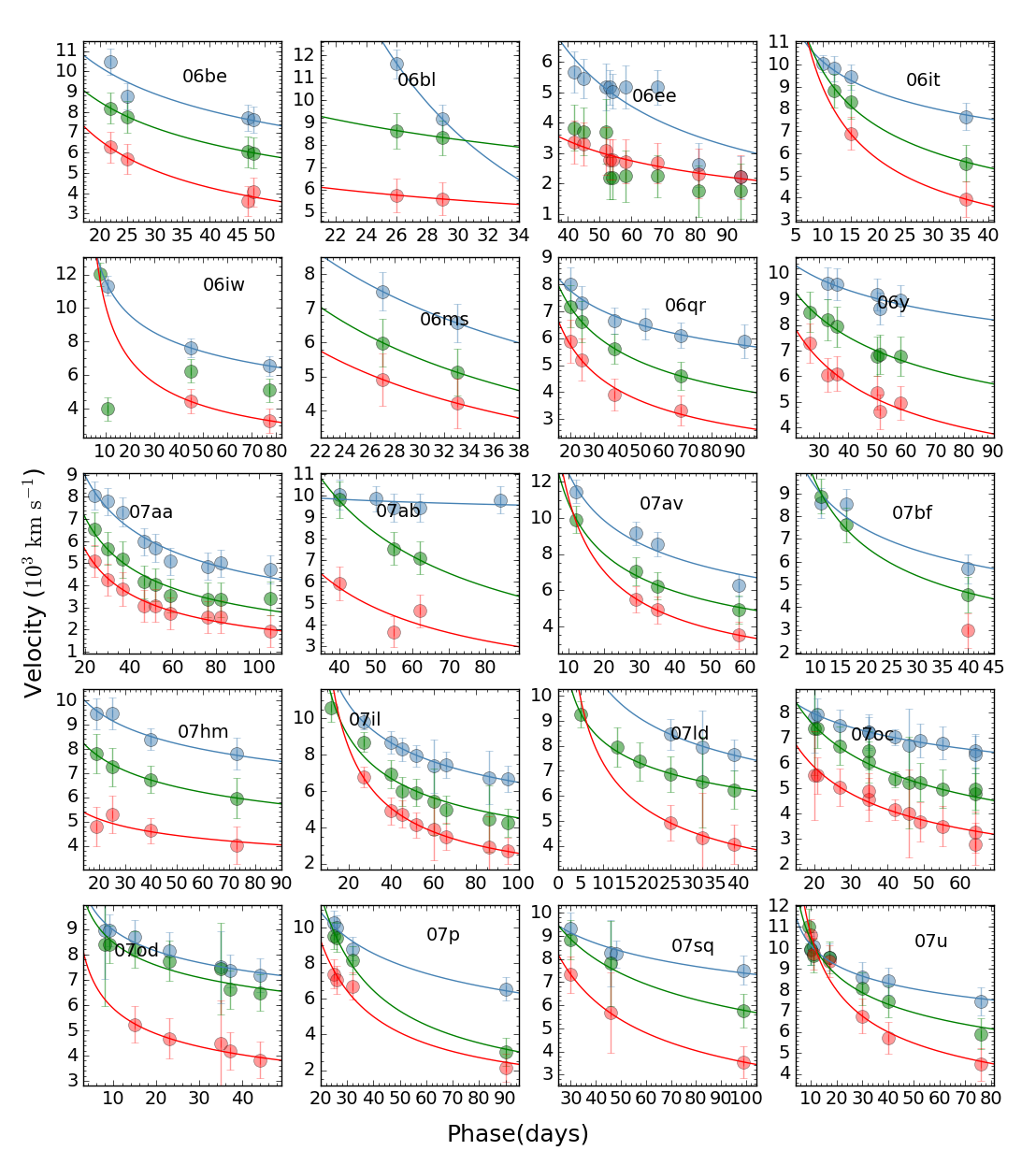}
	\label{fig:gplot_3}
\end{figure*}

\addtocounter{figure}{-1} 

\begin{figure*}
	\includegraphics[width=2.2\columnwidth]{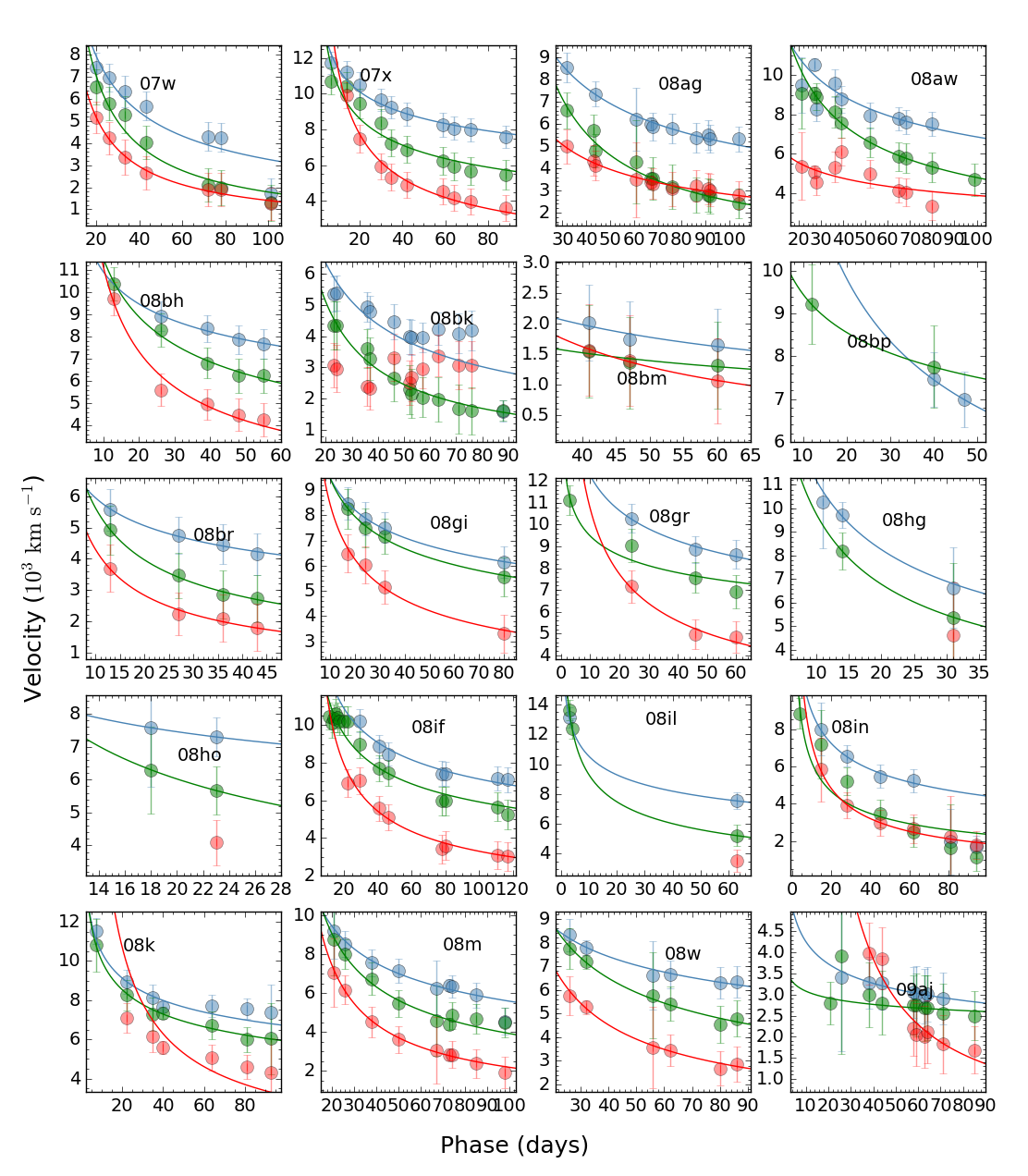}
	\label{gplot_4}
\end{figure*}

\addtocounter{figure}{-1} 

\begin{figure*}
	\addtocounter{figure}{4}
	\includegraphics[width=2.2\columnwidth]{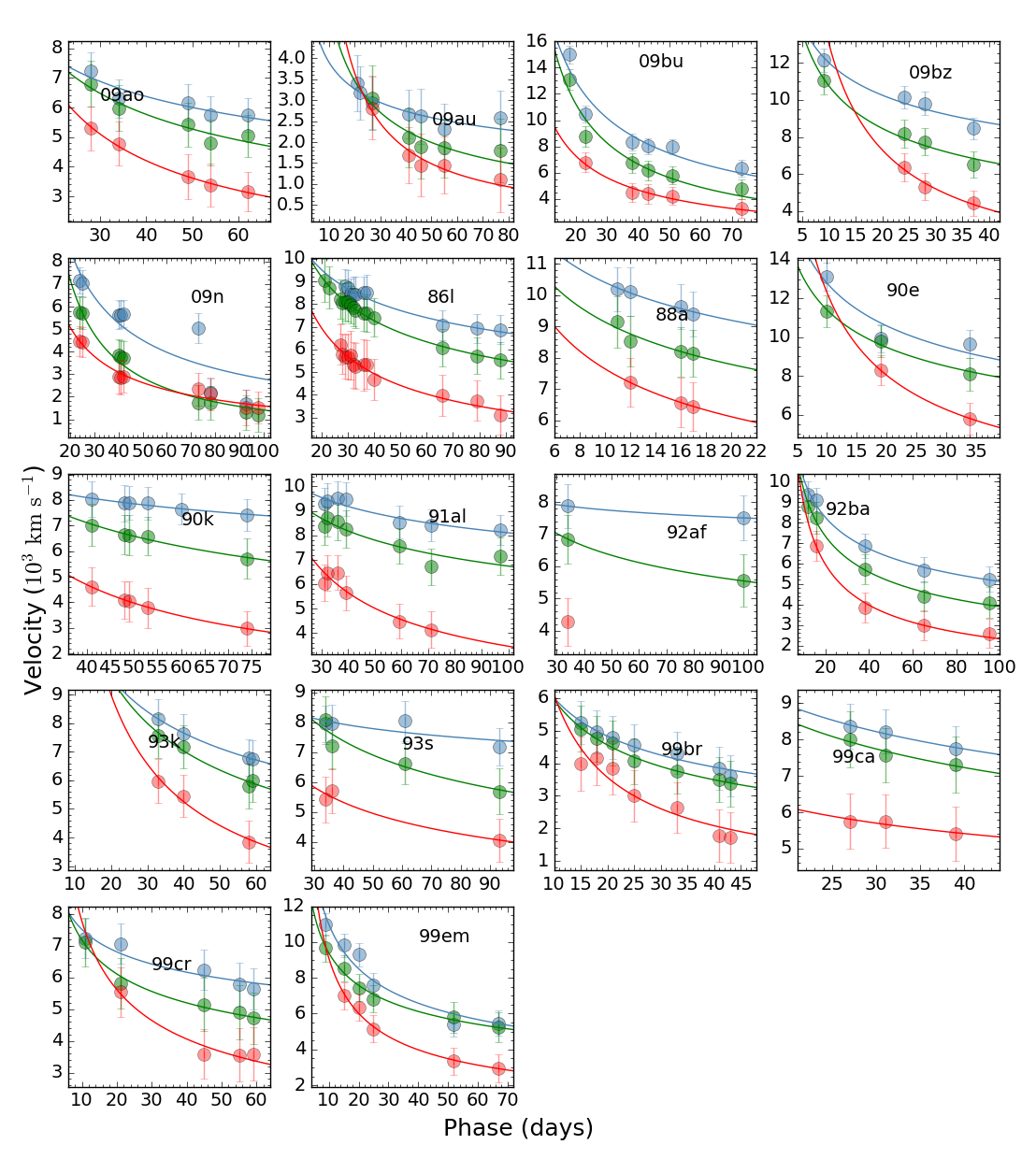}
	\caption{Same as Figure \ref{fig:velocity_show_fit} and Figure \ref{fig:velocityshow_compareobject}, but for the SN II sample in \citet{2017ApJ...850...89G}.}
	\label{fig:gutierrez_powerlaw_fit}
\end{figure*}

\tiny
\setlength{\tabcolsep}{3pt}
\onecolumn
\twocolumn

\onecolumn

\begin{longtable}{lllllllll}
	\caption{SN II sample}   \\
	\label{table:snlog}  \\ 
	\hline
	SN            & RA          & DEC          & Host galaxy                  & Redshift & $E(B-V)_{mw}$ & Last non-detection & First discovery & Explosion date    \\
	\hline	
	\endfirsthead
	
	\multicolumn{9}{c}{SN II sample continue} \\
	\hline	
	SN            & RA          & DEC          & Host galaxy                  & Redshift & $E(B-V)_{mw}$ & Last non-detection & Fist discovery & Explosion date    \\
	\hline	
	\endhead	
	
	\hline
	\endfoot
	
	\hline
	\multicolumn{9}{l}{a:Explosion date estimated by midpoints between the last non-detection and the first discovery.} \\
	\multicolumn{9}{l}{b:Explosion date estimated by matching spectral templates via the SNID.} \\
	\multicolumn{9}{l}{c:First discovery was used as explosion date.} \\
	\multicolumn{9}{l}{d:\citep{2019MNRAS.490.2799D}; e:\citep{2016MNRAS.459.3939V}; f:\citep{2016ApJ...820...33R}; g:\citep{2014ApJ...787..139D}.} \\
	\multicolumn{9}{l}{h:\citep{2015MNRAS.450.2373B}; i:\citep{2017NatPh..13..510Y}; j:\citep{2018MNRAS.478.3776D}; k:\citep{2021MNRAS.503.3472B}.} \\	
	\endlastfoot
	
	ASASSN-14kg   & 01:44:38.38 & 35:48:20.45  & CGCG 521-075                 & 0.014477 & 0.0395 & 56972.38           & 56978.36        & 56969.5(3)$^e$    \\
	ASASSN-14ma   & 23:55:09.13 & 10:12:54.20  & SDSS J235509.00+101252.9     & 0.0137   & 0.087  &                    & 57004.27        & 56994.5(2)$^b$    \\
	ASASSN-15fi   & 16:31:48.80 & 20:24:38.50  & MRK 884                      & 0.017182 & 0.0492 &                    & 57098.63        & 57098.5$^c$  \\
	Gaia15acw     & 08:49:20.24 & 36:42:57.20  & NGC 2668                     & 0.025    & 0.0291 &                    & 57003.33        & 56996.6(2.2)$^b$  \\
	iPTF13dkz     & 01:36:11.58 & 33:37:01.40  & $\cdots$                          & 0.016    & 0.0388 & 56547.5            & 56549.43        & 56547.93(0.45)$f$ \\
	MASTER\_J2103 & 21:03:18.46 & -14:03:51.70 & $\cdots$                          & 0.025    & 0.0407 &                    & 56966.5        & 56952(1.5)$^b$    \\
	MASTER\_J2333 & 23:33:28.09 & -01:58:47.00 & MCG +00-60-06                & 0.017    & 0.0354 &                    & 57300.5        & 57289.7$^b$  \\
	PS15cwo       & 03:11:16.73 & 01:11:34.50  & KUG 0308+010                 & 0.02723  & 0.0893 &                    & 57341.40        & 57334.6(0.2)$^b$  \\
\end{longtable}

\begin{longtable}{lllllll}
	\caption{Journal of Spectroscopic Observations of our SN II sample.}   \\
	\label{table:spec_log}  \\ 
	\hline
	SN                & Obs. date & JD           & Phase$^a$ (Days)              & Instrument & Resolution(\AA $\ \rm pix^{-1}$)  & Wavelength range( \AA )\\
	\hline	
	\endfirsthead
	
	\multicolumn{7}{c}{Journal of Spectroscopic Observations of our SN II sample (continue)} \\
	\hline	
	SN                & Obs. date & JD           & Phase (Days)               & Instrument & Resolution(\AA $\ \rm pix^{-1}$)  & Wavelength range(\AA)  \\
	\hline	
	\endhead	
	
	\hline
	\endfoot
	
	\hline
	\multicolumn{7}{l}{a:Phase relative to the estimated explosion date estimated in Section 2.} \\
	\endlastfoot
	
	ASASSN-14kg        & 20141123 & 2456985.18 & +15.2               & BFOSC  & 4.42 & 3800-8800  \\
	ASASSN-14ma        & 20141214 & 2457006.98 & +12.0d               & YFOSC  & 2.93 & 3500-8900   \\
	ASASSN-15fi        & 20150324 & 2457106.24 & \textgreater{}7.2d  & BFOSC  & 4.42 & 3800-8800   \\
	Gaia15acw         & 20141214 & 2457006.25 & +9.1                & YFOSC  & 7.91 & 3900-8300   \\
	iPTF13dkz         & 20131011 & 2456576.21 & +27.8               & BFOSC  & 4.42 & 4000-8700   \\
	MASTER\_J2103 & 20141108 & 2456970.98 & +18.5               & YFOSC  & 2.93 & 3500-8900   \\
	MASTER\_J2333 & 20151013 & 2457309.00 & +18.8               & BFOSC  & 4.42 & 3800-8800   \\
	MASTER\_J2333 & 20151103 & 2457330.01  & +39.8               & BFOSC  & 18.3 & 3800-8800   \\
\end{longtable}

\begin{longtable}{lllllllll}
	\caption{Line velocities and pseudo equivalent widths of H$\alpha$, H$\beta$ and Fe~{\sc ii} 5169. }   \\
	\label{table:spec_info}  \\ 
	\hline
	SN          & phase               & $v_{\rm H\alpha} (\sigma )^a$            & $v_{\rm H\beta}(\sigma)$            & $v_{\rm Fe II 5169}(\sigma)$       & $\rm pew_{obs}^{H\alpha}(\sigma)$ & |$\rm pew_{emi}^{H\alpha}(\sigma)$| & $\rm pew_{obs}^{H\beta}(\sigma)$ & $\rm pew_{obs}^{Fe II 5169}(\sigma)$ \\
	\hline	
	\endfirsthead
	
	\multicolumn{9}{c}{Line velocities and pseudo equivalent widths of H$\alpha$, H$\beta$ and Fe~{\sc ii} 5169. (continue)} \\
	\hline	
	SN          & phase               & $v_{\rm H\alpha}(\sigma)$            & $v_{\rm H\beta}(\sigma)$            & $v_{\rm Fe II 5169}(\sigma)$       & $\rm pew_{obs}^{H\alpha}(\sigma)$ & |$\rm pew_{emi}^{H\alpha}(\sigma)$| & $\rm pew_{obs}^{H\beta}(\sigma)$ & $\rm pew_{obs}^{Fe II 5169}(\sigma)$ \\
	\hline	
	\endhead	
	
	\hline
	\endfoot
	\hline
	\multicolumn{9}{l}{The unit of velocities is $\rm km\ s^{-1}$ and the unit of pew is \AA} \\
	\endlastfoot
	
	ASASSN-14kg    & +15.2               & 14062(612)     & $\cdots$            & $\cdots$           & 3.88(0.46)   & 96.24(0.09)    & 3.71(0.05)   & $\cdots$              \\
	ASASSN-14ma    & +12.0               & 9942(377)      & 8166(790)      & $\cdots$           & 22.58(1.54)  & 175.33(0.97)   & 44.27(1.10)  & $\cdots$              \\
	Gaia15acw     & +9.1                & 10176(589)     & $\cdots$            & $\cdots$           & 0.26(0.16)   & 60.39(0.27)    & $\cdots$          & $\cdots$              \\
	iPTF13dkz     & +27.8               & 9596(876)      & 6919(470)      & 5788(757)     & 50.93(0.40)  & 227.31(0.15)   & 55.8(0.01)   & 18.14(0.65)      \\
	MASTER\_J2103 & +18.5               & 8747(458)      & 7844(665)      & $\cdots$           & 3.13(0.06)   & 96.48(0.58)    & 24.83(0.45)  & $\cdots$              \\
	MASTER\_J2333 & +18.8               & 11839(349)     & 10155(583)     & 5030(1609)    & 1.45(0.14)   & 177.32(0.06)   & 22.64(0.23)  & 14.16(0.02)      \\
	MASTER\_J2333 & +39.8               & 14958(1343)    & $\cdots$            & $\cdots$           & 2.41(0.78)   & 284.73(1.03)   & $\cdots$          & $\cdots$              \\
	PS15cwo       & +13.1               & $\cdots$            & 9662(2023)     & $\cdots$           & $\cdots$          & $\cdots$            & 16.85(0.01)  & $\cdots$              \\
\end{longtable}

\twocolumn

\begin{table*}
	\centering
	\caption{Velocity at 50 days after explosion and power-law exponent of our sample and well-studied sample in literature.}
	\begin{tabular}{lllllll}
		\hline
		name      & $v_{{\rm H}\alpha}^{50}(\sigma)$      & $n_{{\rm H}\alpha}(\sigma)$           & $v_{{\rm H}\beta}^{50}(\sigma)$      & $n_{{\rm H}\beta}(\sigma)$           & $v_{\rm Fe}^{50}(\sigma)$      & $n_{\rm Fe}(\sigma)$           \\ 
		& ${\rm km\ s^{-1}}$ &   & $\rm km\ s^{-1}$ &   &  $\rm km\ s^{-1}$  \\
		\hline
		PS15cwo   & 8850(203)  & -0.218(0.069) & 7798(269)  & -0.153(0.049) & 5185(644)  & -0.264(0.345) \\
		SN2011az  & $\cdots$        & $\cdots$           & 3356(1373) & -0.861(0.403) & $\cdots$        & $\cdots$           \\
		SN2011bi  & 8396(385)  & -0.201(0.138) & 6922(812)  & -0.513(0.271) & 5439(449)  & -0.255(0.214) \\
		SN2011fu  & $\cdots$        & $\cdots$           & 7118(1555) & -0.176(0.145) & $\cdots$        & $\cdots$           \\
		SN2012A   & 5649(137)  & -0.267(0.048) & 4558(236)  & -0.433(0.081) & 2813(229)  & -0.608(0.128) \\
		SN2012fs  & $\cdots$        & $\cdots$           & 4251(489)  & -0.629(0.097) & $\cdots$        & $\cdots$           \\
\hline  
	\end{tabular}
	\label{tab:v50_n}
\end{table*}

\onecolumn
\begin{longtable}{lllllll}
	\caption{Velocity at 50 days after explosion and power-law exponent of SNe II in \citet{2017ApJ...850...89G}}   \\
	\label{table:gutierrez_v50_n}  \\ 
	\hline
	name      & $v_{{\rm H}\alpha}^{50}(\sigma)$      & $n_{{\rm H}\alpha}(\sigma)$           & $v_{{\rm H}\beta}^{50}(\sigma)$      & $n_{{\rm H}\beta}(\sigma)$           & $v_{\rm Fe}^{50}(\sigma)$      & $n_{\rm Fe}(\sigma)$           \\ 
	& ${\rm km\ s^{-1}}$ &   & $\rm km\ s^{-1}$ &   &  $\rm km\ s^{-1}$  \\
	\hline	
	\endfirsthead
	
	\multicolumn{7}{c}{Velocity at 50 days after explosion and power-law exponent of SNe II in \citet{2017ApJ...850...89G} (continue)} \\
	\hline	
	name      & $v_{{\rm H}\alpha}^{50}(\sigma)$      & $n_{{\rm H}\alpha}(\sigma)$           & $v_{{\rm H}\beta}^{50}(\sigma)$      & $n_{{\rm H}\beta}(\sigma)$           & $v_{\rm Fe}^{50}(\sigma)$      & $n_{\rm Fe}(\sigma)$           \\	
	& ${\rm km\ s^{-1}}$ &   & $\rm km\ s^{-1}$ &   &  $\rm km\ s^{-1}$  \\	
	\hline	
	\endhead	
	
	\hline
	\endfoot
	
	\hline
	\endlastfoot
	
	02fa & 7663(275)  & -0.233(0.103) & 6379(368)  & -0.278(0.159) & 4151(373)  & -0.447(0.231) \\
	02gd & 4146(214)  & -0.213(0.096) & 3430(277)  & -0.422(0.089) & 2460(267)  & -0.476(0.189) \\
	02gw & 6707(213)  & -0.425(0.051) & 5326(260)  & -0.526(0.076) & 3647(251)  & -0.598(0.105) \\
	02hj & 7948(273)  & -0.178(0.065) & 6470(323)  & -0.378(0.091) & 4202(324)  & -0.315(0.141) \\
	02hx & 8087(228)  & -0.194(0.058) & 6464(271)  & -0.267(0.085) & 3821(282)  & -0.288(0.152) \\
	02ig & 8859(470)  & -0.083(0.178) & 7921(1069) & -0.232(0.160) & $\cdots$   & $\cdots$      \\
	03b  & 6284(238)  & -0.341(0.087) & $\cdots$   & $\cdots$      & 3522(344)  & -0.367(0.256) \\
	03bl & 3998(266)  & -0.865(0.184) & 387(484)   & -0.295(0.082) & 2122(468)  & -1.129(0.708) \\
\end{longtable}

\twocolumn
\begin{table*}
	\centering
	\caption{pEW of $\rm H\alpha$ at 50 days after explosion}
	\begin{tabular}{llll}
		\hline
		SN        & $\rm H\alpha_{obs}^{50}(\sigma)$        & $\rm H\alpha_{emi}^{50}(\sigma)$         & |$\rm H\alpha_{o/e}^{50}$|  \\
		\hline
		PS15cwo   & 32.87(0.16)  & -219.65(0.40) & 0.14   \\
		SN2011bi  & 41.41(2.07)  & -186.57(0.53) & 0.24   \\
		SN2012ec  & 58.93(1.71)  & -157.90(1.29) & 0.38   \\
		SN2013gd  & 24.57(0.01)  & -114.51(0.07) & 0.35   \\
		SN2014G   & 21.87(1.22)  & -205.86(0.34) & 0.11   \\
		SN2014cx  & 72.57(0.21)  & -183.46(0.68) & 0.40   \\
		SN2014cy  & 31.01(0.07)  & -82.63(0.02)  & 0.38   \\
		SN2014dq  & 52.89(0.23)  & -184.36(0.21) & 0.29   \\
		SN2015V   & 67.15(0.54)  & -185.65(0.74) & 0.37   \\
		SN2015cz  & 32.61(0.02)  & -137.40(1.15) & 0.24   \\
		SN2016jft & 29.55(0.19)  & -178.35(0.03) & 0.17   \\
		SN2016jfu & 41.56(0.15)  & -127.64(0.91) & 0.33   \\
		SN2017aub & 41.90(0.63)  & -112.56(0.25) & 0.37   \\
		SN2018aoq & 61.09(0.78)  & -115.59(0.40) & 0.56   \\
		SN2012fc  & 54.95(0.25)  & -327.85(0.67) & 0.17   \\
		SN2013af  & 97.99(0.04)  & -178.25(0.39) & 0.55   \\
		SN2014A   & 57.56(0.29)  & -96.20(0.12)  & 0.60   \\
		SN2014B   & 44.88(1.14)  & -224.62(0.24) & 0.20   \\
		SN2015W   & 47.55(0.13)  & -226.44(0.76) & 0.21   \\
		SN2016aip & 60.60(1.76)  & -200.19(2.16) & 0.30   \\
		SN2016jby & 21.19(0.06)  & -123.93(0.18) & 0.17   \\
		SN2012A   & 44.87 (0.13) & -204.44(0.38) & 0.22    \\
		\hline
	\end{tabular}
	\label{tab:pew50}
\end{table*}


\bsp	
\label{lastpage}

\end{document}